\documentclass[aps,reprint,nofootinbib,preprintnumbers,twocolumn,floatfix,superscriptaddress,prx,showpacs]{revtex4-1}
\usepackage{physics}
\usepackage{natbib}
\usepackage{amsmath,amssymb}
\usepackage{graphicx}
\usepackage[dvipsnames]{xcolor}
\usepackage[caption=false]{subfig}
\usepackage{float}
\usepackage[pdftex]{hyperref} 
\usepackage{qcircuit}
\usepackage{wasysym}
\usepackage{bm}

\definecolor{orange}{cmyk}{0,0.6,1.2,0.1}
\definecolor{dblue}{rgb}{0,0.4,0.8}
\definecolor{purple2}{cmyk}{0.5,0.8,0,0}

\begin{document}
	\title{One-way quantum repeater based on near-deterministic photon-emitter interfaces}
	\author{Johannes Borregaard}
	\affiliation{QuTech and Kavli Institute of Nanoscience, Delft University of Technology, 2628 CJ Delft, The Netherlands}
	\affiliation{QMATH, Department of Mathematical Sciences, University of Copenhagen, DK-2100 Copenhagen \O, Denmark}
	\author{Hannes Pichler}
	\affiliation{ITAMP, Harvard-Smithsonian Center for Astrophysics, Cambridge, MA 02138, USA}
	\affiliation{Department of Physics, Harvard University, Cambridge, MA 02138, USA}
	\author{Tim Schr{\"o}der}
	\affiliation{Department of Physics, Humboldt-Universit{\"a}t, 12489 Berlin, Germany}
	\affiliation{Center for Hybrid Quantum Networks (Hy-Q), The Niels Bohr Institute, University of Copenhagen, DK-2100 Copenhagen \O, Denmark}
	\author{Mikhail D. Lukin}
	\affiliation{Department of Physics, Harvard University, Cambridge, MA 02138, USA}
	\author{Peter  Lodahl}
	\affiliation{Center for Hybrid Quantum Networks (Hy-Q), The Niels Bohr Institute, University of Copenhagen, DK-2100 Copenhagen \O, Denmark}
	\author{Anders S. S\o rensen}
	\affiliation{Center for Hybrid Quantum Networks (Hy-Q), The Niels Bohr Institute, University of Copenhagen, DK-2100 Copenhagen \O, Denmark}
	\date{\today}
	\date{\today}
	
	\begin{abstract}
We propose a novel one-way quantum repeater architecture based on photonic tree-cluster states. Encoding a qubit in a photonic tree-cluster protects the information from transmission loss and enables long-range quantum communication through a chain of repeater stations. As opposed to conventional  approaches that are limited by the two-way communication  time, the overall transmission rate of the current quantum repeater protocol is determined by the local processing time enabling very high  communication rates. We further  show that such a repeater can be constructed with as little as two stationary qubits and one quantum emitter per repeater station, which significantly increases the experimental feasibility. We discuss potential implementations with diamond defect centers and semiconductor quantum dots efficiently coupled to photonic nanostructures and outline how such systems may be integrated into repeater stations. 
	\end{abstract}
	\maketitle

\section{Introduction}
Encoding information in quantum systems is the fundamental principle of quantum information technologies, ranging from quantum computers~\cite{Ladd2010} to unconditionally secure communication~\cite{Gisin2002}. Quantum networks constitute an important element for implementing such technologies in a scalable fashion~\cite{Kimble2008}. The exact requirements and applications of  large-scale quantum networks constitute an active research area~\cite{Wehner2018}. One of the key challenges for constructing large scale quantum networks is to faithfully transmit quantum information over long distances which is challenging due to transmission loss. 

Quantum repeaters have been proposed as a means to overcome transmission loss by exploiting quantum correlations to extend the transmission length of quantum information~\cite{Briegel1998,Munro2015,Muralidharan2016}. The conventional quantum repeater architecture relies on heralded quantum entanglement distribution, which necessitates  long-lived quantum memories and two-way communication between sender and receiver~\cite{Munro2015}. The need for heralding limits the communication rate at which quantum information can be distributed and requires long-lived quantum memories with efficient light-matter coupling~\cite{Seri2017,Borregaard2018}. To overcome these limitations, one-way and all-photonic quantum repeaters have been proposed~\cite{Fowler2010,Munro2012,Muralidharan2014,Glaudell2016,Azuma2015,Ewert2016,Lee2018}. One-way repeaters use multi-photon encoding and quantum error-correcting codes to protect the quantum information from both loss and operational errors. In this way, quantum information can be transmitted from one repeater station to the next without the need for pre-established entangled links. For these reasons, in principle, one-way repeaters can significantly boost the distribution rate~\cite{Muralidharan2016} without the need for a long-lived quantum memory for key applications such as long-distance quantum key distribution~\cite{Scarani2009}.  
An outstanding challenge involving the physical implementation of one-way quantum repeaters is how to efficiently generate the multi-qubit error-correcting codes and how to perform error correction. This usually requires many high-fidelity two-qubit operations and considerable amounts of auxiliary qubits at each repeater station~\cite{Muralidharan2014,Lee2018,Ewert2016,Ewert2017}.
\begin{figure*} [t]
\centering
\includegraphics[width=0.90\textwidth]{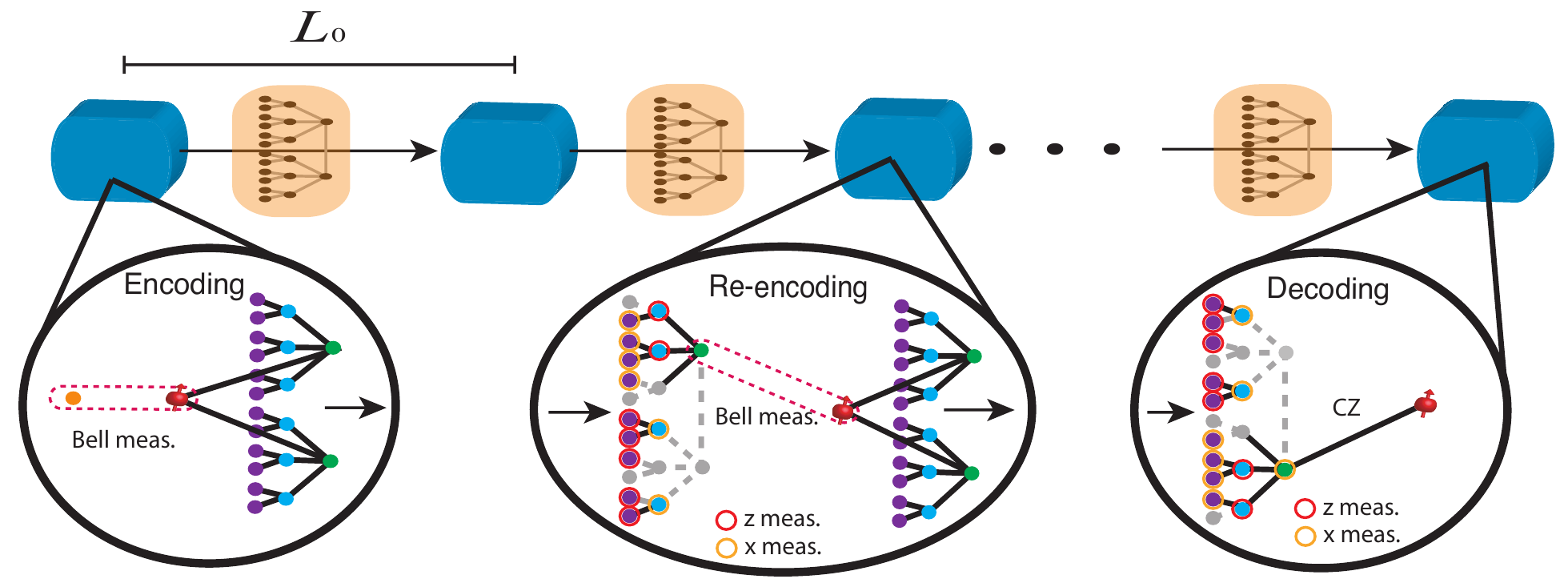}
\caption{Sketch of a one-way quantum repeater with photonic tree-clusters. A message qubit (orange dot) is encoded through a Bell measurement (dashed box) with the root spin qubit of a photonic tree-cluster. An example with a $[2,3,2]$-tree is shown where dots correspond to photonic qubits and solid lines indicate correlations. The encoded qubit (depicted in the orange boxes) is sent a distance $L_0$ to the next repeater station. At the next repeater station, the qubit information is re-encoded through a Bell measurement between one of the 1st-level photonic qubits and the root spin qubit of a new tree. The remaining qubits of the incoming tree are measured with single qubit  $z$ (red circle) or $x$ (orange circle) measurements. The re-encoding can succeed despite multiple photons being lost in transmission (gray dots and dashed lines). At the end station, the message qubit can be transferred to e.g. a receiving spin qubit by means of a controlled phase (CZ) gate.}
\label{fig:figure1}
	\end{figure*}	
	
In this Article, we propose a novel one-way quantum repeater architecture that can be implemented with as little as two memory qubits and one quantum emitter per repeater station. Our approach is based on photonic tree-cluster states~\cite{Varnava2006}, which are used to encode a message qubit to be transmitted to the next repeater station (see Fig.~\ref{fig:figure1}). Photonic tree-clusters have previously been considered as photonic memories to ensure efficient entanglement swapping in all-optical quantum repeaters~\cite{Azuma2015,Pant2017}. Such all-optical approaches generate multiple photonic tree-clusters at each repeater station potentially requiring kilometer long delay lines and millions of single-photon sources per station~\cite{Pant2017}. Our approach circumvents this significant overhead by using strongly coupled quantum emitters with build-in nonlinearity.
Specifically,  in our approach, the photonic tree-clusters required for the repeater can be generated with two memory qubits and one single photon emitter per repeater station using repeated photon emissions~\cite{Buterakos2017}. In addition, correction of losses only require  a single Bell measurement independent of the size of the tree-encoding. This constitutes a significant reduction in overhead as compared to, e.g., one-way quantum repeaters based on the quantum parity encoding~\cite{Munro2012, Muralidharan2014,Ewert2016},
which requires a number of two-qubit operations that scales linearly with the size of the encoding corresponding to 
 hundreds of memory qubits per repeater station~\cite{Muralidharan2016}. In comparison, 
  the current approach can be implementated with only two spin systems per repeater station as we outline below. We also discuss possible experimental implementations of our protocol based on state-of-the-art solid-state quantum emitters in nanophotonic structures in order to lay out a realistic path towards high bit-rate, long-range quantum communication. Importantly, many of the required parameters for our protocol are not far from current state-of-the art performances, which together with the significant resource reduction compared to previous one-way protocols cements the experimental feasibility of our approach. 
  
\section {Quantum repeater protocol}
The basic operation of the repeater is shown in Fig.~\ref{fig:figure1}. At each node, a multi-photon entangled state is generated and used to encode and transmit a message qubit to the next repeater station. Crucially, even if some of the photons are lost, the repeater can decode the logical qubit and re-encode it, thereby correcting for photon loss before transmitting the message to the next station.  
\subsection{Tree-cluster states}
In this work, we consider one way quantum repeaters based on using tree-cluster states as error correcting codes. Such tree-cluster states are illustrated in Fig.~\ref{fig:figure2}. We characterize the tree by a branching vector $\bar{t}=[b_0, b_1, \ldots, b_d] $, which specifies the connectivity of the tree as one moves from the root vertex (top node in Fig.~\ref{fig:figure2}(a)) through the $d$ levels of the tree. Tree-cluster states are obtained by associating a qubit with each vertex (see Fig.~\ref{fig:figure1}). Moreover, one further associates a stabilizer operator $K_i=\sigma^x_i \bigotimes_{j\in \mathcal{N}(i)} \sigma^z_j$, that acts nontrivially on the vertex $i$ and its neighbors $\mathcal{N}(i)$. The tree-cluster state is the unique eigenstate with eigenvalue +1 of all stabilizer operators $K_i$.

As a specific, illustrative example, let us consider a [2,2]-tree cluster state (see Fig.~\ref{fig:figure2}(a)). One can easily check that this 7-qubit state is given by 
\begin{align}
\ket{\psi}=\frac{1}{2^{3/2}}(&\ket{0}_{\text{s}}(\ket{0}\ket{+,+}+
\ket{1}\ket{-,-})^{\otimes 2 }\nonumber\\
+&\ket{1}_{\text{s}}(\ket{0}\ket{+,+}-\ket{1}\ket{-,-})^{\otimes 2 })
\end{align}
Here, we have defined the states $\ket{\pm}=(\ket{0}\pm\ket{1})/\sqrt{2}$, with $\ket{0}$ and $\ket{1}$ being the basis states of the qubits. Below we consider an implementation where the root qubit is represented by a stationary two-level spin system (spin qubits are denoted with subscript s) while the rest of the tree cluster state is represented by photons.  


\subsection{Encoding the logical qubit}
\begin{figure} [t]
\centering
\includegraphics[width=0.49\textwidth]{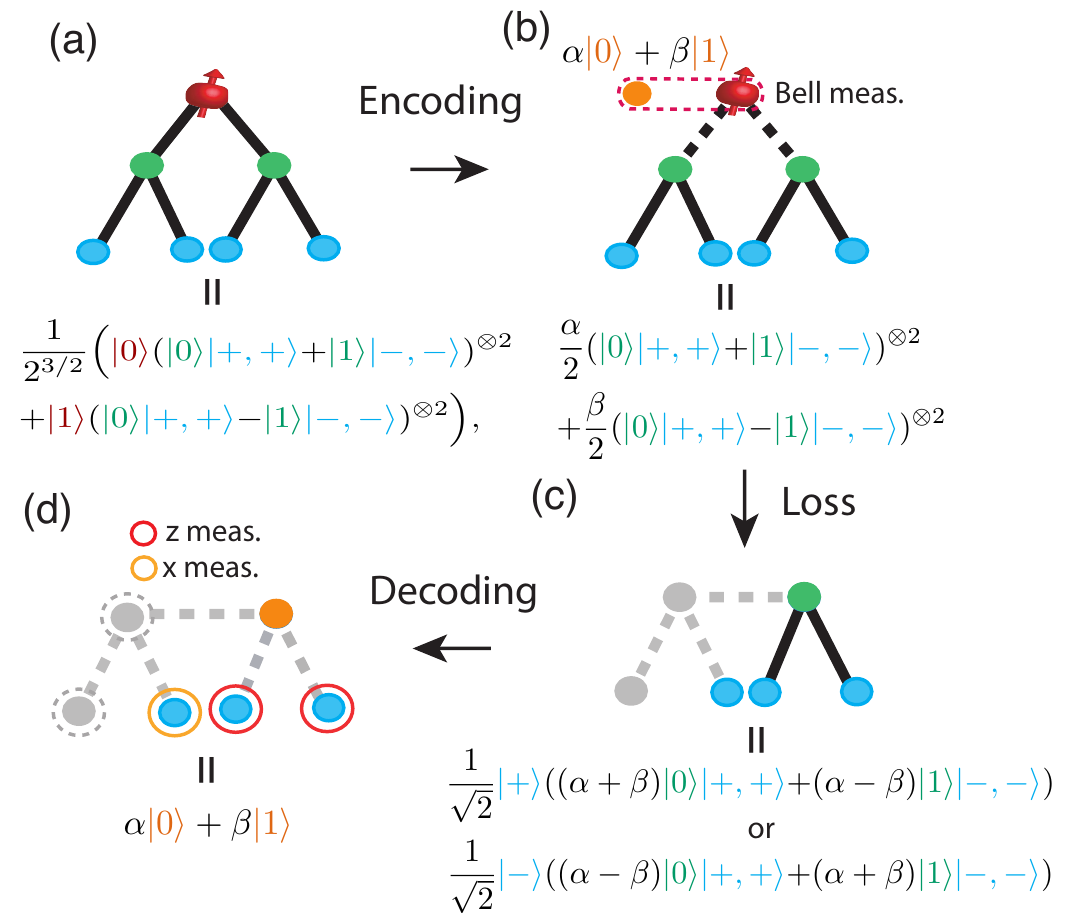}
\caption{Sketch of the principle behind counter-factual error-correction in the tree-cluster encoding. A [2,2] tree is considered for simplicity (a). A qubit $\alpha\ket{0}+\beta\ket{1}$ is encoded through a Bell measurement between this qubit and the root spin qubit (b). Loss of qubits (gray vertices) from one branch (c) still allows retrieving the encoded information by measuring at least one remaining 2nd-level qubit in the $x$ basis and the second level qubits of the intact branch in the $z$ basis (d) despite the fact that we do not get any information from (attempted) measurements on the lost photons (dashed circles).}
\label{fig:figure2}
	\end{figure}

Consider the situation, where the message qubit, $\alpha\ket{0}_{\text{s}}+\beta\ket{1}_{\text{s}}$, is initially prepared in a second stationary qubit. To send this message qubit from the first station to the next repeater, one has to encode it into the state of the photons. This can be achieved by a simple teleportation process, which can be realized by a 2-qubit Bell measurement of the stationary qubits, i.e. the spin that stored the message qubit and the stationary root qubit of the tree-cluster state (see Fig.~\ref{fig:figure2}(b))

For the above example of a $[2,2]$-tree this prepares the state of the photons in  
\begin{eqnarray} \label{eq:22enc}
 \ket{\Psi}&=&\tfrac{(1+x_1)\alpha+(1-x_1)\beta}{4}\left(\ket{0,+,+}+\ket{1,-,-}\right)^{\otimes 2} \nonumber \\
&& +x_2 \tfrac{(1-x_1)\alpha+(1+x_1)\beta}{4}\left(\ket{0,+,+}-\ket{1,-,-}\right)^{\otimes 2},
\end{eqnarray}
where $x_1=\pm 1$ and $x_2=\pm 1$, depending on the four possible outcomes of the Bell measurement. The values of $x_1$ and $x_2$ are not important since they can eventually be corrected in the decoding step. In Fig.~\ref{fig:figure2}, we assume for concreteness that we have obtained the values $x_1=x_2=1$. Note that the quantum information of the message qubit is stored in a non-local form in the photonic degrees of freedom, and can not be retrieved by observing, e.g., only a single photon. 

\subsection{Photon loss and recovery of the message qubit}
After this encoding step, the photons are transmitted to the next repeater station. The specific encoding protects against transmission loss such that the effective transmission probability of the message qubit is significantly increased compared to the bare transmission of a single photon. 

To illustrate the basic mechanism, we again consider the example of the [2,2] encoding in Eq.~(\ref{eq:22enc}). Already in this simple encoding, one can tolerate the loss of up to two photons in one of the two branches (see Fig.~\ref{fig:figure2}(c)). To see this, it is instructive to consider how the state can be recovered and the quantum information retrieved. As a first step in the recovering process, one measures all the qubits in one of the two branches (in Fig.~\ref{fig:figure2}, we assume that the left branch is to be measured). Specifically, the first level qubit is measured in the $z$ basis and the second level qubits are measured in the $x$ basis. Note that the corresponding measurement outcomes are perfectly correlated, such that only two sets of outcomes for the three measurements are possible. This is crucial, as it allows to infer the outcome from each of the three measurements, even if two of those qubits are lost. 
This measurements projects the state of the qubits in the remaining branch (right branch in Fig~\ref{fig:figure2}(c)) into the state
\begin{align}
\tfrac{(1+x_1)\alpha+(1-x_1)\beta}{2\sqrt{2}}(\ket{0,+,+}+\ket{1,-,-})\nonumber \\+x_2 x_3 \tfrac{(1-x_1)\alpha+(1+x_1)\beta}{2\sqrt{2}}(\ket{0,+,+}-\ket{1,-,-})
\end{align}
where $x_3=\pm 1$, depending on the measurement outcome. This branch now contains the entire encoded quantum information. It can be retrieved by measuring the two second level qubits in the $z$ basis (see Fig.~\ref{fig:figure2}(d)). Simple algebra shows that this prepares the remaining first level qubit in the state $\alpha\ket{0}+\beta\ket{1}$ (up to known Pauli corrections that only depend on the obtained measurement outcomes, $x_i$).  This simple analysis shows that the retrieval of the message qubit from a [2,2]-tree is possible as long as one branch is not corrupted and not more than two qubits of the other branch are lost, illustrating the basic principle allowing for correction of photon loss. Increasing the tree depth (length of $\bar{t}$) and the number of branches increases the robustness of the encoding by the same principle~\cite{Varnava2006}. 

\subsection{Re-encoding and repetition}
The goal of the repeater station is to re-encode the retrieved qubit in a new tree. This can be achieved in complete analogy to the encoding of the message qubit at the sending station: first a new tree-cluster of photons is generated, with a stationary spin serving as the root qubit (see Sec.~\ref{sec:schemeI}), followed by a Bell measurement between the message qubit and the root qubit. 

Above we described how to recover the qubit into photonic qubit in the highest level of the tree. In the re-encoding procedure the goal is instead to perform a Bell measurement between the encoded qubit and the root qubit of a new tree. In analogy with the procedure above this re-encoding simply requires a Bell measurement between one of the 1st level photons of the encoded tree-cluster and the root qubit along with measurement of all other qubits in the same bases as above. Note that the order of the measurements is not important in the above recovery scheme. In practice, this allows us to re-encode the quantum information at each repeater station without prior knowledge about which qubit was lost. Specifically, one can first attempt a Bell measurement between one of the first level qubits and the root qubit of the new tree. If this measurement is successful, one can teleport the encoded quantum information into the new tree by measuring all connected qubits in the $z$ basis. Some of these measurements may turn out to be unsuccessful because the qubits were lost in transmission. In these cases, the corresponding measurement outcome is inferred through measurements on qubits in the next level of the corresponding branch, in complete analogy to the example above. If the first Bell measurement is unsuccessful itself (because the corresponding photon was lost in transmission), then the value of a $z$ measurement can be inferred instead (via measurement of the next level qubits), and a Bell measurement can be attempted with another first level qubit. In order not to perturb the root qubit of the new tree in a failed attempt of a Bell measurement, special care must be given to the implementation of the measurement, as described below. Specifically, the message qubit should first be transferred to an auxiliary spin qubit by means of a spin-photon controlled-phase gate (CZ-gate) and then encoded into the new tree-cluster with a deterministic Bell measurement between the auxiliary spin qubit and the root spin qubit (see Fig.~\ref{fig:figure3} below).  

The re-encoding and transmission continues down the repeater chain until the encoded message qubit arrives at the end node. There the message qubit can be either transferred to a stationary spin in a similar fashion as in the repetition step (see Fig. \ref{fig:figure1}), or directly measured (without first transferring the information to a receiving spin qubit) by appropriate measurements of the photons of the encoded tree.



\section{Experimental implementation}
The key requirements for an implementation of the above protocol are the ability to generate tree-cluster states of photons, realize Bell-measurements between stationary spins and photonic qubits, and perform measurements of the photons in the  $x$ and $z$ basis. 
\subsection{Photonic tree generation} \label{sec:schemeI}
We propose to generate the photonic tree cluster states using a light matter interface illustrated in Fig.~\ref{fig:figure3}. It consists of stationary memory spins and one spin which is coupled to the light field. The latter is used to generate photons by selectively coupling a ground state $\ket{1}_{\text{s}}$ to an excited $\ket{e}_{\text{s}}$ via the optical field in a one-sided cavity. We are considering a time-bin representation of the photonic qubits. In this representation, the presence of a single photon in one of two non-overlapping spatiotemporal modes represents the two qubit states $\ket{0}$ and $\ket{1}$. The main reason for using this time-bin representation is that it allows to detect errors stemming from photon loss. This could also be obtained with a polarization representation but the time-bin representation is better suited for transmittance through optical fibers, which typically disturb the polarization state.

Recent work~\cite{Buterakos2017} showed that sequential excitation of the quantum emitter, together with controlled phase gates between the emitter and the memory spins allows to deterministically generate an arbitrary photonic tree-cluster state. In particular, a tree-cluster state of depth $d$ only requires $d-1$ memory spin systems while the number of necessary spin-spin CZ-gates scales polynomially with the number of branches at each level~\cite{Buterakos2017}. In what follows, we show that tree-cluster states of depth 3 are sufficient for transmission distances up to 1000 km (assuming telecom frequencies) and consequently only two memory spin systems and a single quantum emitter are necessary for the generation of such states.

\begin{figure} [t]
\centering
\subfloat{\includegraphics[width=0.26\textwidth]{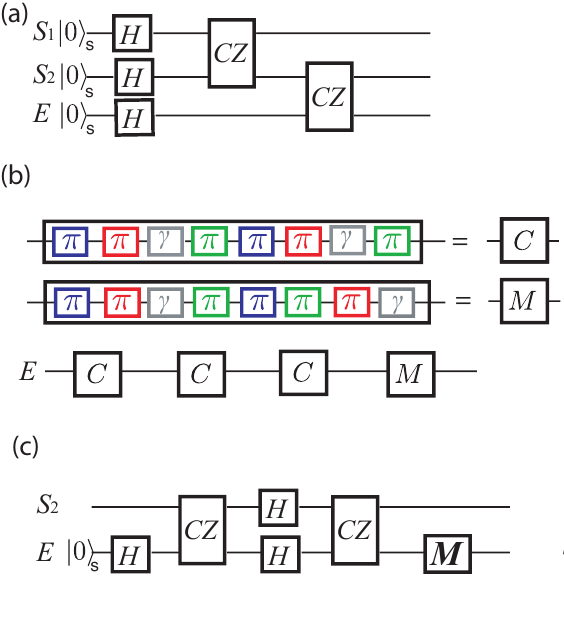}}
\subfloat{\includegraphics[width=0.24\textwidth]{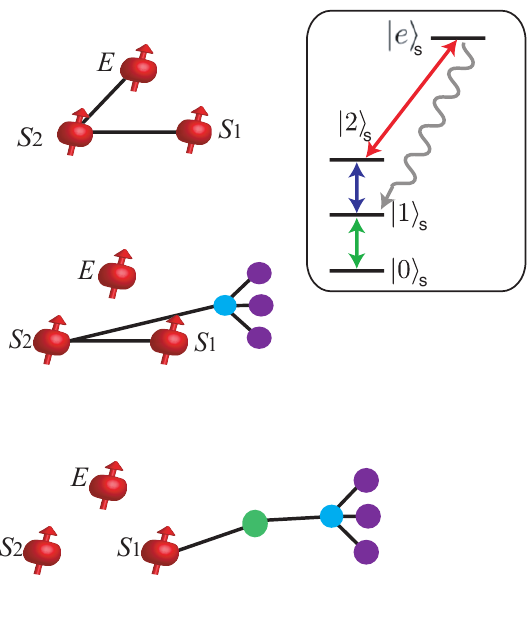}}
\caption{Generation of photonic tree-cluster states. Two memory spin systems ($S_1,S_2$) and a quantum emitter ($E$) are all prepared in state $\ket{+}_{\text{s}}$ and CZ-gates entangle the spins as shown in (a). In the second step (b), all 3rd level photons (purple dots) of one subbranch of the tree-cluster are emitted (operation $C$) followed by the emission of the corresponding 2nd level photon (blue dot), which detaches the emitter from the preliminary tree-cluster (operation $M$). Operations $M$ and $C$ can be implemented using a sequence of $\pi$ pulses on various transitions in the emitter as indicated by the color code in (b) referring to the color of the transitions in the emitter level diagram shown in the inset. Here driven transitions are indicated by solid lines, whereas decay ($\gamma$) is represented by the wiggle line. To emit more 2nd level qubits, $E$ is again prepared in $\ket{+}_{\text{s}}$  followed by a CZ-gate with $S_2$ and step (b) is repeated. In step (c), the state of the memory spin is first swapped to the emitter and then emitted as the 1st level photon of the branch. Here the emission of the 1st level photon is achieved by an operation $\bm{M}$, which is similar to $M$ except that the red $\pi$ pulse and subsequent decay is replaced by a weak driving of the $\ket{2}_{\text{s}}-\ket{e}_{\text{s}}$ transition to ensure that the photons have a narrow bandwidth. The steps are then repeated until the entire tree has been emitted.  }
\label{fig:figure4}
	\end{figure}

The generation of a tree-cluster state with depth 3 is sketched in Fig.~\ref{fig:figure4}. In the first step (a), CZ-gates are applied between the two memory spins and between the emitter and one memory spin. The spin of the quantum emitter acts as a second level qubit in the tree and all third level photonic qubits of the subbranch are emitted through repeated excitation followed by spontaneous emission as shown in  Fig.~\ref{fig:figure4}(b). The second level qubit is then mapped to a photonic qubit using the auxiliary level of the emitter, which also detaches the emitter from the preliminary tree-cluster state. This step is repeated until all 2nd and 3rd level photons of the branch have been emitted. The spin of the second memory qubit is then first swapped to the spin of the emitter by means of two CZ gates and subsequently swapped to a photonic qubit (Fig.~\ref{fig:figure4}(c)). This completes the emission of one branch of the photonic tree-cluster state and the procedure can then be repeated to output the entire state. We note that this generation procedure, in principle, require additional Hadamard gates on the 3rd level photons to fit the stabilizer description introduced earlier. This is, however, not necessary since it is sufficient to simply rotate the measurement basis of these photons in the re-encoding and decoding steps.   

A drawback of this generation scheme is that the photons will be emitted such that the 1st-level qubit of a branch is emitted last. As previously described, the presence/absence of a 1st-level photon determines the measurement basis of the corresponding branch at a repeater station. It is therefore necessary to delay the photons of each branch to enable measuring the 1st-level qubit first. The length of this delay will depend on the emission rate of the emitters and the number of photons per branch, but is generally modest and implementable in optical fiber delays. We will discuss this issue in more detail below.

\subsection{Bell measurement}
As described above, the re-encoding at the repeater stations requires a successful Bell measurement between one of the 1st level qubits in the encoded tree-cluster and the root qubit of the new tree-cluster. It is crucial, that this Bell measurement is designed with a built-in error detection: if a Bell measurement is attempted with a lost 1st level qubit, the measurement should abort without perturbing the root qubit. Otherwise, a new tree-cluster has to be generated after each failed attempt.

\begin{figure} [t]
\centering
\includegraphics[width=0.49\textwidth]{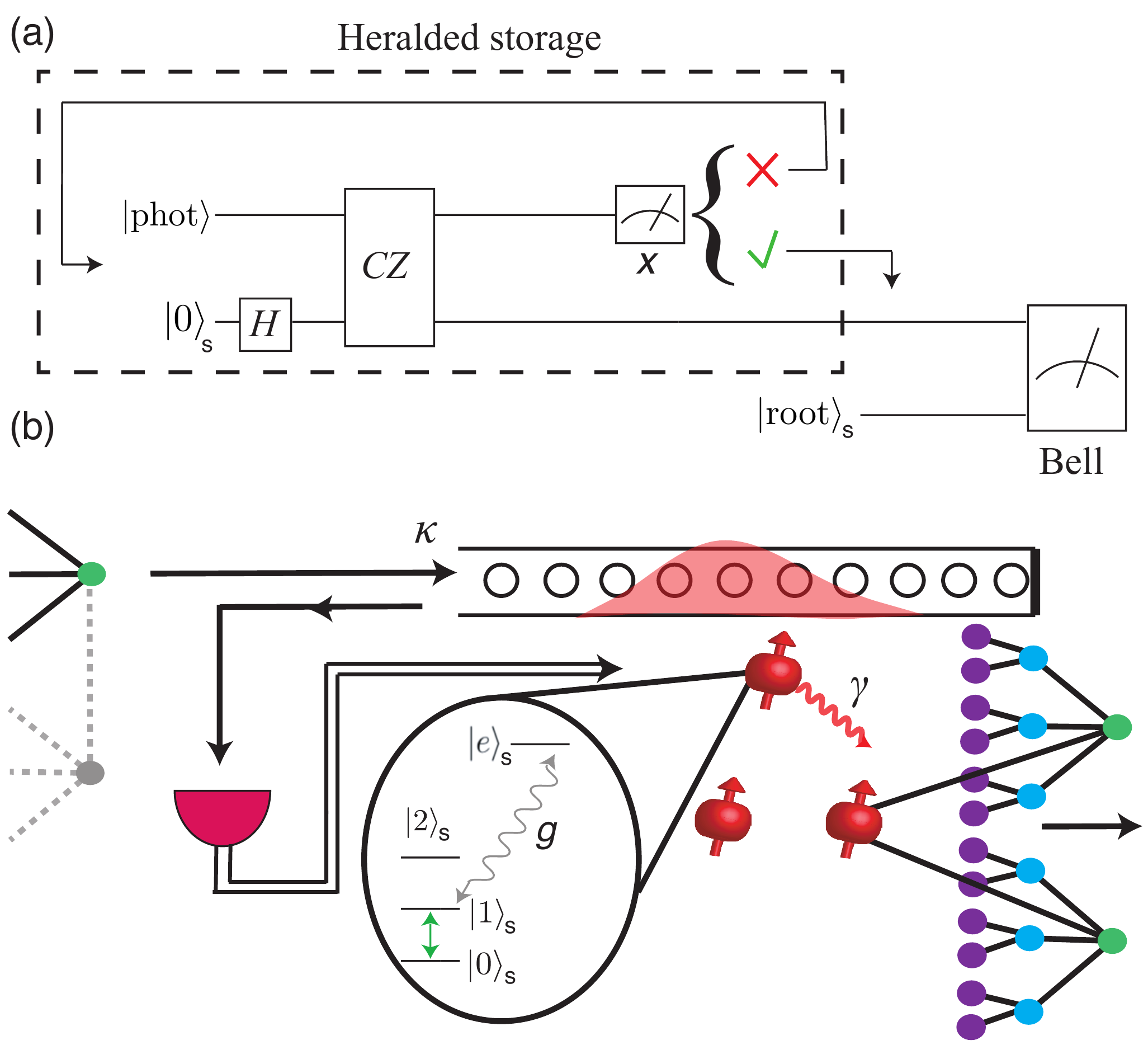}
\caption{The basic elements of the re-encoding operation at the repeater stations. (a) The re-encoding is performed in a loss tolerant manner by first performing a heralded storage of the message qubit in an auxiliary memory spin (dashed box). This is obtained through a spin-photon controlled phase (CZ) gate with a 1st level photon of the encoded tree. The transfer is heralded by the detection of the photon in the $x$-basis. If unsuccessful (photon loss), the auxiliary spin is initialized and the operation is tried again with another first level photon. Once the storage is successful, a deterministic Bell measurement is performed between the auxiliary spin and the root spin qubit of the new tree-cluster thereby completing the re-encoding operation. (b) The experimental implementation of the re-encoding operation. The spin-photon CZ gate is performed by reflecting the photonic qubit off a one-sided cavity (shown as a photonic crystal cavity) strongly coupled to a quantum emitter as described in the text. The parameters $g,\gamma$, and $\kappa$ are the single photon Rabi frequency of the optical transition, the spontaneous emission rate of the emitter, and the decay rate of the cavity field, respectively. The same physical setup is used to generate the tree-cluster state requiring an extra spin qubit and one auxiliary level ($\ket{2}_{\text{s}}$) in the quantum emitter (see Fig.~\ref{fig:figure4}).}
\label{fig:figure3}
	\end{figure}

The setup required to generate the tree-cluster states (Fig.~\ref{fig:figure4}) conveniently also allows for such an operation. While one of the stationary memory qubits represents the root qubit, the spin coupled to a one-sided cavity is used for heralded storage of the message qubit through a spin-photon CZ-gate (see below)~\cite{Reiserer2014,Tiecke2014,Kalb2015,Sun2018}. Importantly, the success of the storage is conditioned on subsequently detecting the photon in the $x$-basis (see Fig.~\ref{fig:figure3}). When a storage attempt is unsuccessful due to the loss of the photonic qubit, the auxiliary spin qubit is simply re-initialized and a new attempt is made with another first level qubit. The root spin qubit of the new tree-cluster is completely unaffected by this.  Once the storage is successful, a Bell measurement between the auxiliary spin-system and the spin-system containing the root qubit of the new tree is performed using deterministic entangling gates between the two spin systems, concluding the re-encoding step. 

To perform the cavity mediated spin-photon CZ-gate, we assume the auxiliary spin to initially be prepared in ground state $\ket{0}_{\text{s}}$. For a time-bin encoded photonic qubit, the early half of the wave packet corresponding to qubit state $\ket{0}$ is first reflected off the cavity. In the ideal limit, the photon will be reflected with a $\pi$-phase shift from the cavity. Now the transformation $\ket{0}_{\text{s}}\to(\ket{0}_{\text{s}}+\ket{1}_{\text{s}})/\sqrt{2}$ is performed on the emitter before the late half of the photon wave packet, corresponding to qubit state $\ket{1}$, is reflected off the cavity. If the auxiliary spin system is in state $\ket{1}_{\text{s}}$ ($\ket{0}_{\text{s}}$), the photon gets reflected without (with) a $\pi$ phase shift in the limit of strong light-matter interaction $C\gg1$, characterized by the cooperativity $C=g^2/(\kappa \gamma)$. Here $g$ is the single photon Rabi frequency of the cavity mediated $\ket{1}\leftrightarrow\ket{e}$ transition, $\kappa$ is the decay rate of the cavity and $\gamma$ is the free space spontaneous emission rate of the excited level. Up to a global phase, this amounts to a CZ-gate between the photonic time-bin qubit and a qubit in the ground states of the auxiliary spin system initially prepared in $(\ket{0}_{\text{s}}+\ket{1}_{\text{s}})/\sqrt{2}$. The details of the gate interaction and main errors are described in the supplemental material~\cite{SM}. The success of the gate is conditioned on subsequently detecting the photon in the $x$-basis, which boosts the fidelity. We find that if the intra-cavity losses are tuned to be on the order of $1/C$ then a spin-photon cooperativity of $C=100$ is sufficient to ensure an error $\lesssim10^{-4}$ and a success probability $\sim0.99\eta_d$, where $\eta_d$ is the efficiency of the photon detection. 

In the above estimate, we have assumed that reflection of photons into the detector from e.g., imperfect mode-matching is negligible. Such events directly translate into an error since they correspond to operation without any spin-photon interaction. Careful engineering of the mode profile and additional filtering of, e.g., uncoupled polarization modes must therefore be employed to suppress such reflections to the desired error-level. We note, however, that a mode-matching efficiency of 99\% is sufficient to ensure an error of $10^{-4}$~\cite{SM}. Furthermore, we have assumed that the frequency width of the 1st level photons is narrow enough to neglect errors from the finite bandwidth of the Purcell enhanced emitter. For $C=100$, this would require the 1st level photons to have a frequency width $\sim\gamma$ to have errors $\lesssim10^{-4}$~\cite{SM}. Weak driving from an auxiliary level $\ket{2}_{\text{s}}$ to $\ket{e}_{\text{s}}$ (see inset in Fig.~\ref{fig:figure4}) allows to tune the emission time of the first level photons to achieve this~\cite{Pichler2017} (operation $\bm{M}$ in Fig.~\ref{fig:figure3}). 

We note that depending on the success of the spin-photon Bell measurement, the measurement basis of the qubits in the corresponding branch of the encoded tree-cluster states must be adjusted. If a 1st-level qubit is lost (detected), the qubits in the corresponding branch should be measured in the $x$ ($z$) basis for the tree-cluster states generated as described above. 



\subsection{Photon measurement} 
With the considered time-bin encoding, measurements of photons in the $z$-basis require only time-resolved detection. Measurements in the $x$-basis on the other hand are more demanding. In particular, a deterministic $x$-basis measurement requires fast optical switching and delay lines. Our analysis shows that GHz optical switching rates will be required to ensure tree-generation rates in the MHz regime (see below). Such switching rates can be exceeded with schemes based on sub-nanosecond phase-control in Mach-Zehnder interferometers (MZI) via electro-optic (EO) modulation. Such integrated devices have been demonstrated for a variety of material platforms~\cite{reed2010,sun2015,ogiso2017,koeber2015,krasnokutska2018,wang2018}. 
Towards a scalable, small footprint implementation we therefore propose an on-chip photonic circuit based on switching via EO modulation in cascaded Mach-Zehnder interferometers as shown in Fig.~\ref{fig:measurements} and discussed in the supplemental material where we also outline an integrated on-chip setup for the repeater stations~\cite{SM}.
\begin{figure} [t]
\centering
\includegraphics[width=0.40\textwidth]{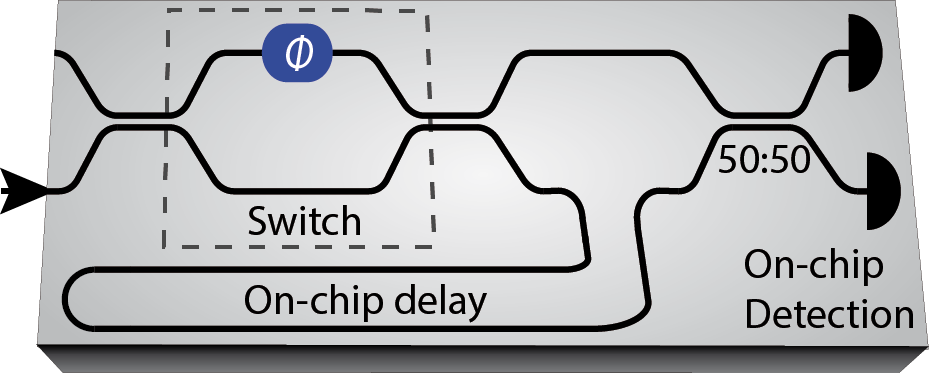}
\caption{Illustration of an on-chip photonic circuit for performing $z$- and $x$-basis measurements with integrated single photon detectors. Mach-Zehnder interferometers with fast switching rate, for example based on electro-optic switching~\cite{reed2010,sun2015,ogiso2017,koeber2015,krasnokutska2018,wang2018}, are assumed to switch fully between the top and bottom path (labeled switch) 
by controlling the relative phase of the two interferometer arms ($\phi$). For a $z$-basis measurement, the photons are guided directly to the detectors for time-resolved recording. For an $x$-basis measurement, the switch guides the first wave package of the time-bin encoding to the delay line and the second to the short arm. For 1st level photons the delay need to be longer and is replaced with a fibre delay line~\cite{SM}. 
}
\label{fig:measurements}
	\end{figure}
	
\subsection{Other experimental requirements}	
\begin{figure*} [t]
\centering
\subfloat{\label{fig:figure5a}\includegraphics[width=0.485\textwidth]{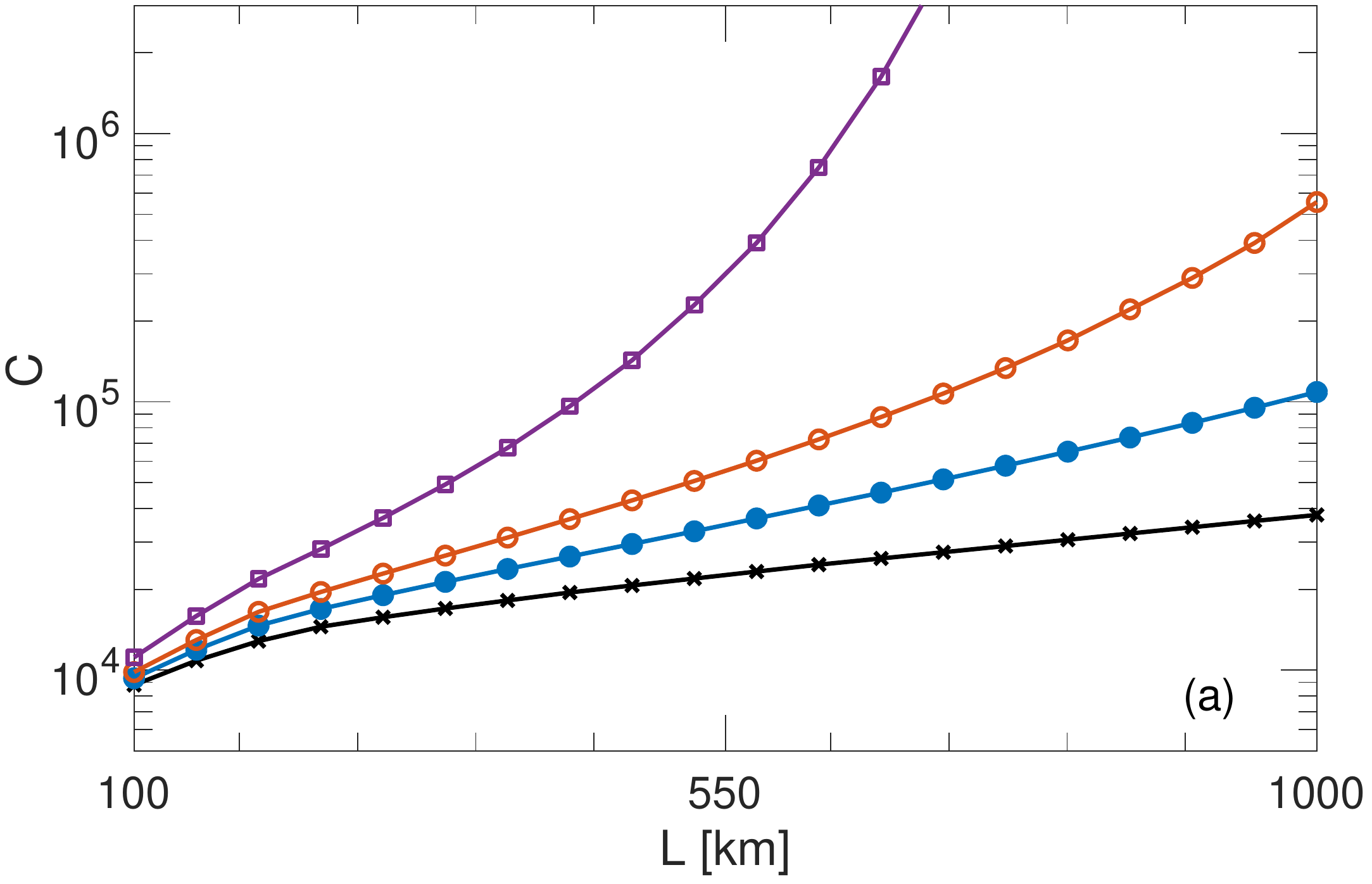}} \subfloat{\label{fig:figure5b}\includegraphics[width=0.48\textwidth]{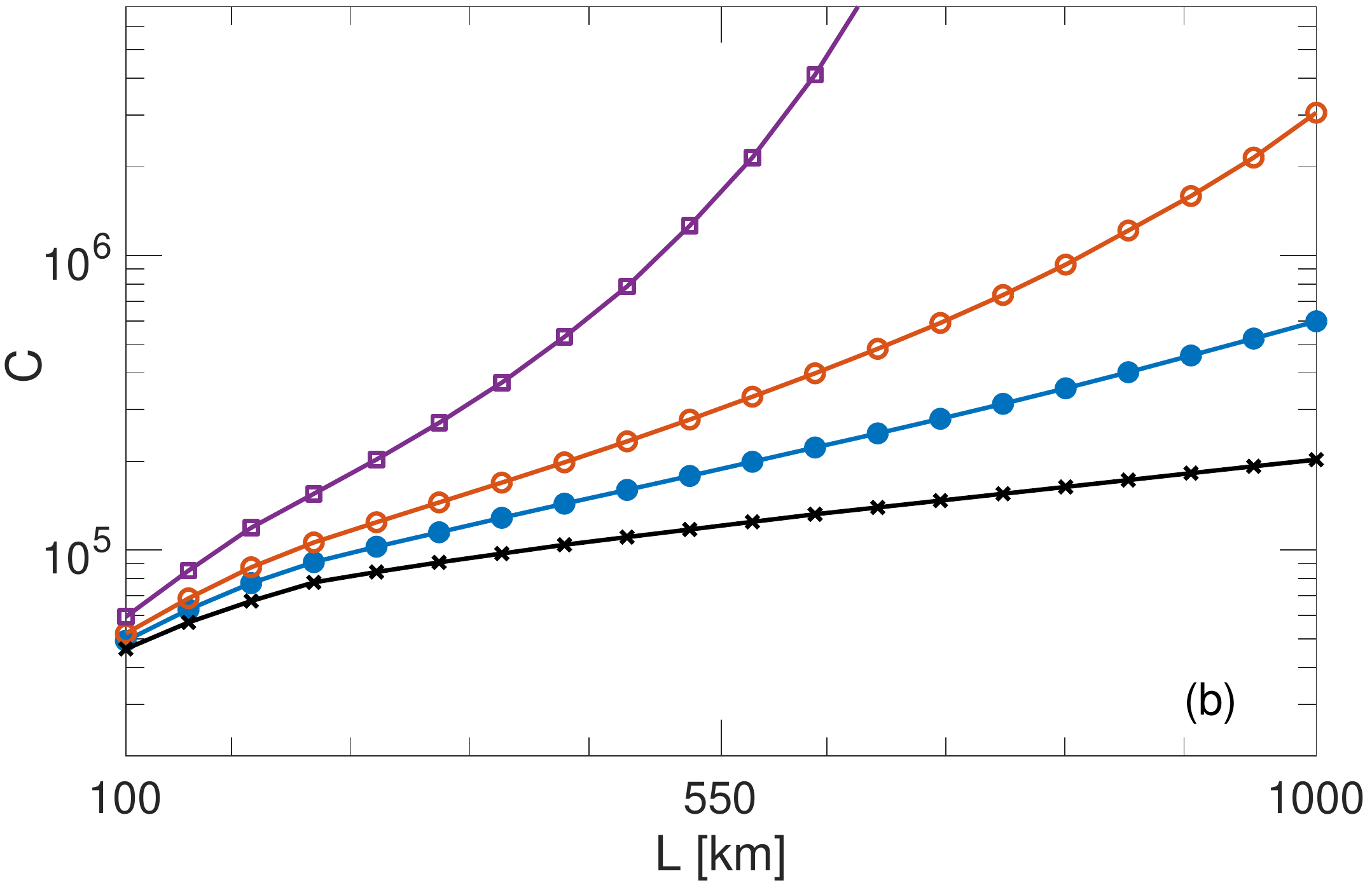}} \\
\subfloat{\label{fig:figure5c}\includegraphics[width=0.48\textwidth]{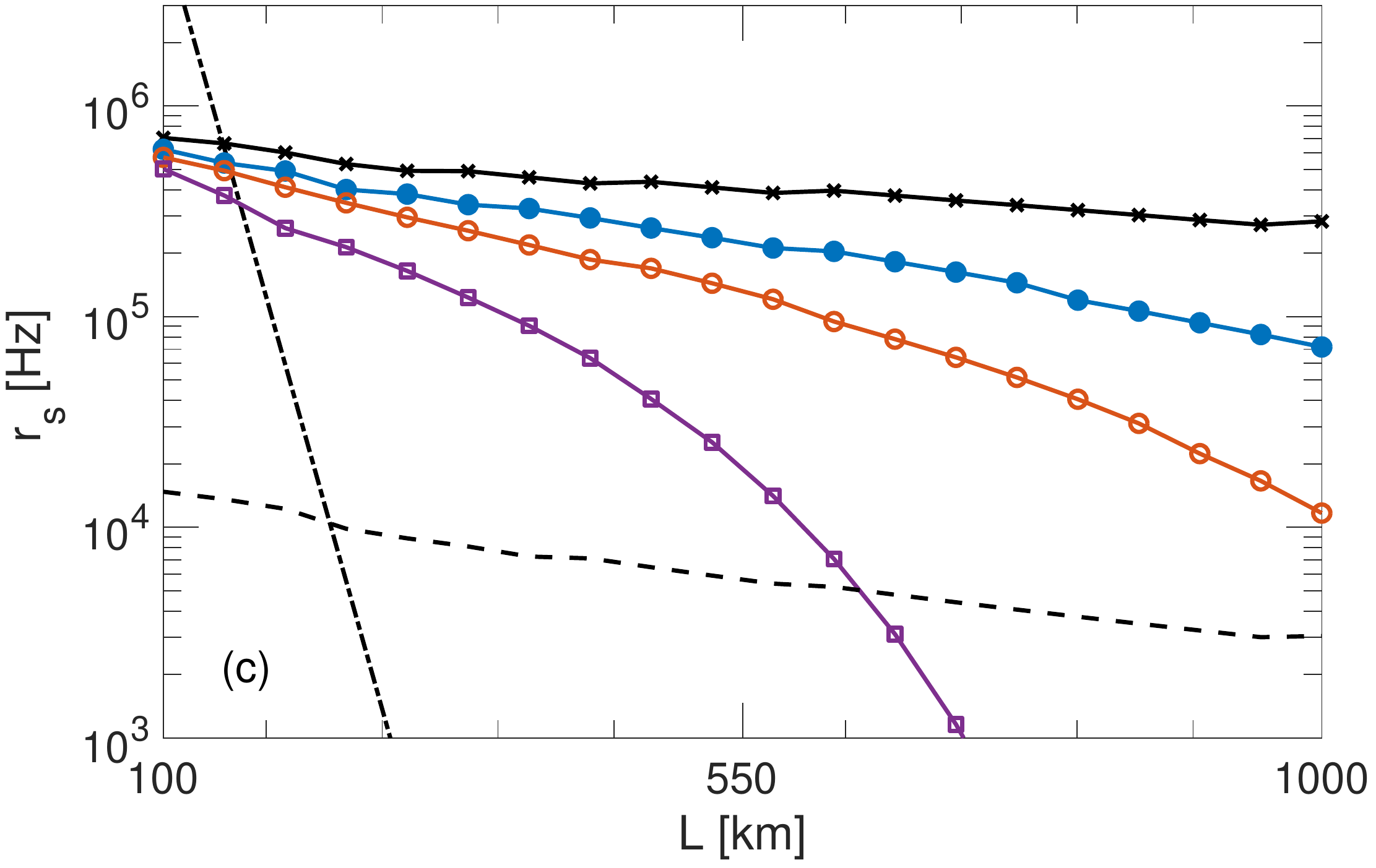}} \subfloat{\label{fig:figure5d}\includegraphics[width=0.475\textwidth]{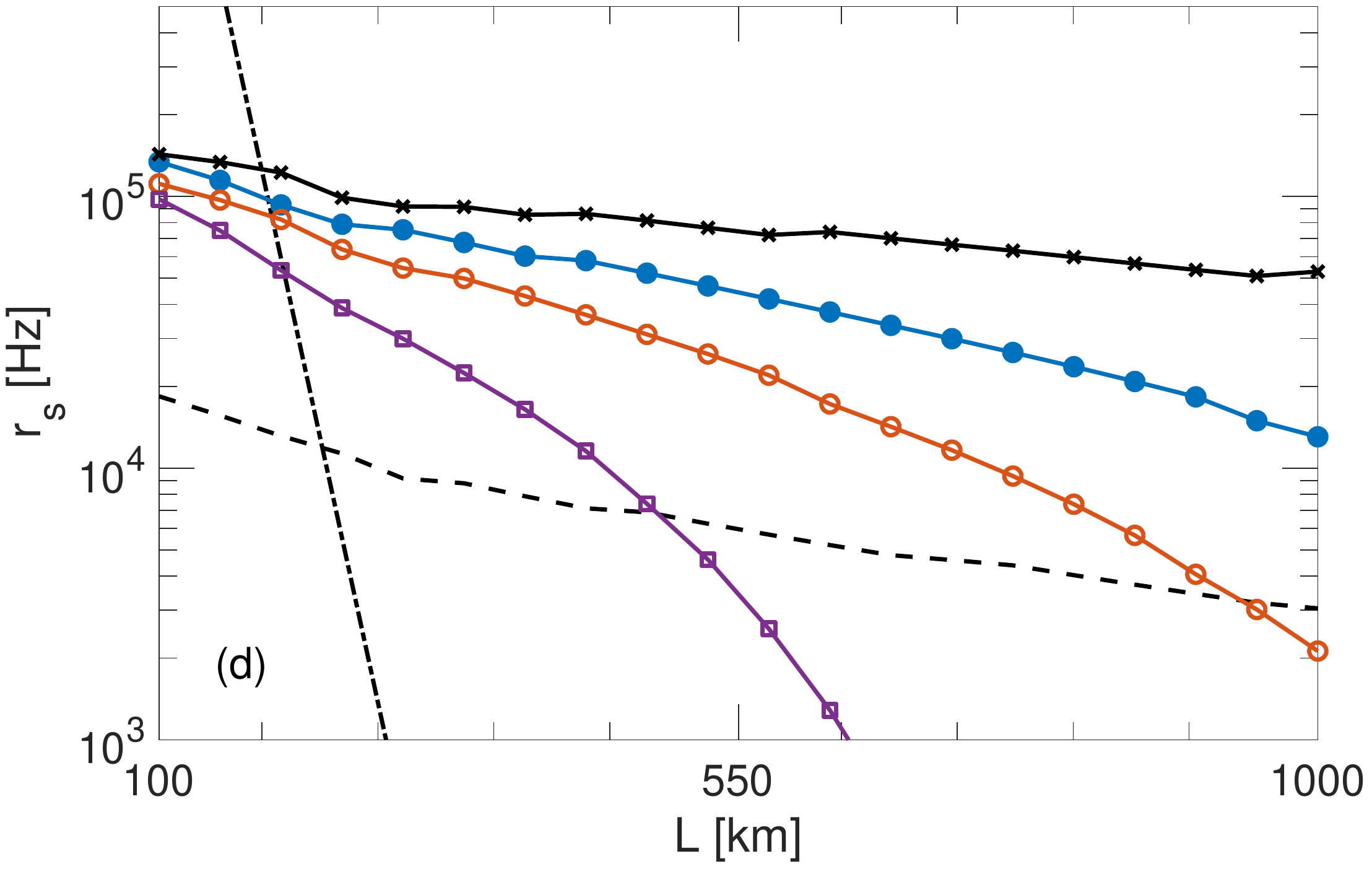}}
\caption{(a)-(b) Minimized cost parameter $C$ as a function of distance $L$ assuming spin entangling gate times of $\tau_{\text{CZ}}=10\tau_{\text{ph}}$ (a) or $\tau_{\text{CZ}}=100\tau_{\text{ph}}$ (b) for tree-cluster state generation. The markers correspond to: $\bm{\times}:\epsilon_r=0.1\permil$, $\textcolor{dblue}{\bullet}:\epsilon_r=0.3\permil$, $\textcolor{orange}{\bm{\circ}}:\epsilon_r=0.5\permil$, and $\textcolor{purple2}{\bm{\square}}:\epsilon_r=0.1\%$. Here $\epsilon_r$ is the error probability of the re-encoding operation at the repeater stations. (c)-(d) Corresponding secret bit rate assuming a photon emission time of $\tau_{\text{ph}}=1$ ns and gate time of $\tau_{\text{CZ}}=10$~ns (c) and $\tau_{\text{CZ}}=100$~ns (d). We have restricted the minimization to trees with $n\leq300$ photons and repeater station spacings $\geq 1$ km. We have assumed a detection efficiency of $\eta_d=95\%$. For comparison, the secret bit rate of a two-way repeater (dashed line) with similar resources (see main text) is also plottet together with the rate of direct transmission (dot-dashed line) assuming a 1 GHz single photon source. }
\label{fig:figure5}
\end{figure*}

So far we have discussed the optical interface required to achieve the successful operation of the repeater. In order to be able to make realistic estimate of the achievable  communication rate, we will now discuss concrete requirement for two specific physical systems, quantum dots and color centers in diamond. One of the practical requirements in reducing the photon loss is a highly efficient coupling for the cavity to an optical fiber, e.g. using tapered optical fibers~\cite{Tiecke2015,Patel2016,Daveau2016}. The collection efficiency ($\beta$-factor) of the emitted photons to a cavity or alternatively a waveguide needs to be high. Quantum dots in waveguides have already demonstrated collection efficiencies of $\beta > 98\%$~\cite{Acari2014}, which is compatible with the efficiency $\eta_d=0.95$ assumed in our resource analysis below (see Fig.~\ref{fig:figure5}). In addition, coupling to photonic nanostructures may also decrease the photon emission time through the Purcell enhancement and photon emission times of $\sim100$ ps are feasible with solid state emitters such as quantum dots~\cite{Liu2018} and diamond color centres~\cite{Zhang2018}.

Finally, spin-spin CZ gates are required both for the tree-generation and the re-encoding operation. Fast spin-spin gates ($\sim 10$ ns) could be performed in stacked quantum dots~\cite{Kim2010, Evans2018} while somewhat slower gates can be performed between electron and nuclear spins for Nitrogen-vacancy (NV) or Silicon-vacancy (SiV) centers in diamond through magnetic dipolar interaction~\cite{Kalb2017}. For the latter, gate times on the order of $100$ ns are feasible with SiV systems using nearby nuclei with strong ($>$1 MHz) hyperfine interactions. Alternatively, fast gates could also be implemented using photon mediated gates between different emitters~\cite{Mahmoodian2016}. This involves two auxiliary spin qubits for parity measurements and can be made error-proof against photon loss errors at the expense of a slight decrease in success probability. For the gate in question~\cite{Mahmoodian2016}, a $\beta$-factor of $\sim 0.99$ would give a success probability of $\sim99\%$ and a heralded error of $\sim0.1\permil$. Such probabilisitic spin-spin gates will, however, decrease the rate of the repeater when used in the re-encoding step at the repeater stations. The reason being that the re-encoding involves the (unprotected) root qubit of the new tree. Considering a distance of 1000 km where $\sim400$ repeater stations are needed (see optimization below), a success probability of $99$\% would result in a rate that is $\sim2$\% of the rate for a deterministic gate. Using probabilistic gates in the tree generation steps is of less concern since heralding techniques can be employed and the majority of the necessary gates will involve the redundant qubits of the tree-encoding, which are somewhat loss-tolerant.  

The details of the time budget for the generation of the tree-cluster states and length of the necessary delay line is detailed in the supplemental material~\cite{SM}. We find that the photonic tree-cluster states can be emitted within $\sim 1$ $\mu$s ($\sim7$ $\mu$s)  assuming Purcell-enhanced photon emission lifetime of about 100~ps~\cite{Liu2018,Zhang2018} and spin-spin CZ gate times of $10$~ns ($100$~ns). The spin qubits need to stay highly coherent for these time scales, which for SiV and NV systems can be achieved using nearby nuclear spins~\cite{Maurer2012} or operating at low temperatures~\cite{Sukachev2017}. For quantum dots, dynamical decoupling~\cite{Bluhm2010} or coupling to a nuclear spin memory~\cite{Gangloff2019} may be employed to increase coherence times motivating further development of such techniques. For the above generation times, delay lines of maximum length $\sim68$ m ($\sim374$ m) at the repeater stations are required ensure the right detection order of the photons (see Sec.~\ref{sec:schemeI}). At telecom frequencies such a delay line would have a transmission above $99\%$ ($\sim98\%$), which can be integrated into the overall detection efficiency $\eta_d$.

\section{Repeater performance}
The above analysis outlined all necessary operations and general hardware considerations of the repeater. Importantly, we have shown that only a single successful Bell measurement is needed at the re-encoding step and that this can be implemented in a loss-tolerant manner using two spin systems. Furthermore, we have outlined how the photonic tree-clusters may be generated requiring in total only two qubit spin systems per repeater station in addition to the quantum emitter. We now proceed by estimating the performance of the repeater in terms of the maximum quantum bit rate for given distances.

The transmission probability of a message qubit through the entire repeater chain will be 
\begin{equation}
p_{\text{trans}}=\eta_{e}^{m+1},
\end{equation}
where $m$ is the number of equally spaced repeater stations between the start and end stations, and $\eta_{e}$ is the transmission probability of the encoded quantum information between repeater stations. The encoded transmission probability depends on the specific tree-encoding, the bare transmission probability $\eta$ of a single photon between repeater stations, and the detection efficiency of the photon detectors $\eta_d$.  Note that in/out coupling efficiency and any frequency conversion efficiency that may be required to transduce to the telecom band can be directly included in $\eta_d$. 

For a tree-cluster encoding with branching vector $\vec{t}=[b_0, b_1,..b_d]$, $\eta_e$ is given by the recursive formula~\cite{Varnava2006}
\begin{equation}
\eta_{e}=\left(\left(1-\mu+\mu R_1\right)^{b_0}-\left(\mu R_1\right)^{b_0}\right)\left(1-\mu+\mu R_2\right)^{b_1},
\end{equation}
where
\begin{equation}
R_k=1-\left(1-(1-\mu)\left(1-\mu+\mu R_{k+2}\right)^{b_{k+1}}\right)^{b_k},
\end{equation}
with $R_{d+1}=0,b_{d+1}=0$ and $\mu=1-\eta\eta_d$. Here, $R_{k}$ is the probability of having a successful indirect $z$ measurement of any given qubit in the $k$'th level of the tree. Consequently, the total probability of a successful $z$ measurement of a $k$'th level qubit (direct or indirect) is $1-\mu+\mu R_{k}$. For a fiber-based implementation, the bare transmission will be $\eta=\text{exp}(-L_0/L_{\text{att}})$, where $L_0$ is the distance between the repeater stations and $L_{\text{att}}=20$ km is the attenuation length of the optical fiber assuming that efficient frequency conversion to the telecom band is implemented.

The relative simplicity of the encoding in tree-cluster states comes with the penalty that it is not able to correct arbitrary errors as opposed to other codes considered for one-way repeaters. It is clear that an error on the qubits participating in the re-encoding Bell-measurement will map into an error on the encoded message qubit. However, there is some robustness against errors due to the large redundancy of information encoded in the tree~\cite{Azuma2015}. As discussed below this leads to an error rate of the encoded qubits, which is only a few times the error rate of the individual qubits. 

To quantify the performance of the tree repeater in the presence of operational errors, we consider the secret bit fraction $f$ of the transmitted qubits, which can be estimated in the asymptotic limit of infinitely long keys assuming perfect classical error correction. Assuming QKD is performed using a six-state variant of the BB84 protocol we have that~\cite{Scarani2009}
\begin{equation}
f=1-h(Q)-Q-(1-Q)h\left(\frac{1-3Q/2}{1-Q}\right),
\end{equation}
where $Q=2\epsilon_{\text{trans}}/3$ is the qubit error rate of the transmitted bits and $h(x)=-x\log_2(x)-(1-x)\log_2(1-x)$ is the binary entropy. We have assumed a (worst-case) scenario where the noise on the transmitted bits are described by a single qubit depolarizing channel of the form
\begin{equation} \label{eq:depol}
\Lambda(\hat{\rho})=(1-\epsilon_{\text{trans}})\hat{\rho} + \frac{\epsilon_{\text{trans}}}{3}\left(\sigma^x\hat{\rho}\sigma^x+\sigma^y\hat{\rho}\sigma^y+\sigma^z\hat{\rho}\sigma^z\right)
\end{equation}
where $\hat{\rho}$ is the density matrix of the (pure) message qubit and $\sigma^{x,y,z}_i$ are the Pauli matrices. The final error probability of the transmitted message qubit is $\epsilon_{\text{trans}}$. For a repeater with $m$ repeater stations, it is estimated as $\epsilon_{\text{trans}}=1-(1-\epsilon_{r})^{m+1}\approx(m+1)\epsilon_{r}$ for $m\epsilon_{r}\ll1$ where $\epsilon_{r}$ is the error probability of the re-encoding step at the repeater stations. Note that $f$ is negative for $Q\gtrsim12.61$\% reflecting that it is no longer possible to extract any secret bits from the transmitted qubits since they are too noisy for privacy amplification. Since the tree-encoding is not able to correct the errors, this will eventually limit the distance to $L\sim0.13L_{0}/\epsilon_r$.

\subsection{Optimization of repeater performance}
In order to asses the performance of the repeater, we perform a numerical optimization of the number of repeater stations ($m$) and the encoding tree ($\vec{t}$) for a given distance and error ($\epsilon_{r}$) to find the highest possible secret bit rate. We assume that the local repetition rate, $r_0$, is set by the emission time of the photonic tree-cluster states. For realistic parameters (see below), this will be determined by the emission time of a photonic qubit ($\tau_{\text{ph}}$) and the gate time ($\tau_{\text{CZ}}$) of spin-spin gates. For a specific tree-encoding ($\vec{t}=[b_0,b_1,\ldots,b_d]$) we estimate the generation time as
\begin{eqnarray}
\frac{1}{r_{0}}&\sim&b_0[100+b_1(1+b_2(1+\cdots b_{d-1}(1+b_d)\cdots))]\tau_{\text{ph}} \nonumber \\
&&+b_0[3+b_1(1+b_2(1+\cdots b_{d-2}(1+b_{d-1})\cdots))]\tau_{\text{CZ}}.
\end{eqnarray}
Note that we assume the emission time of the first level photons to be $100\tau_{\text{ph}}$ to have errors $\sim10^{-4}$ in the scattering gate of the re-encoding step (see above). In addition, three spin-spin entangling gates are needed for the creation of the first level qubits. We then seek to minimize the (dimensionless) cost parameter 
\begin{equation} \label{eq:cost}
C=\frac{1}{r_0fp_{\text{trans}}}\times\frac{ m L_{\text{att}}}{\tau_{\text{ph}}L},
\end{equation}
where the first factor is the inverse secret key while the second factor includes the extra cost of adding repeater station.  The inverse cost parameter can be viewed as the secret key rate per repeater station per attenuation length for a given total distance $L$.  In the optimization, we enforce a maximum of the number of photons in the encoding of $n=300$ and require that the repeater stations are never placed closer than 1 km apart. The results of the optimizations are shown in Fig.~\ref{fig:figure5}.

It is clear that as the operational errors increase, the repeater performs worse since the tree-encoding is not fault-tolerant with respect to depolarizing errors. Nonetheless, for $\epsilon_{r}\lesssim0.1\%$, it is still possible to reach high secret bit rates since the repetition rate is determined by the local repetition rate, which can be in the MHz regime. Specifically, with a photon emission time of $\tau_{\text{ph}}=1$ ns and gate time of $\tau_{\text{CZ}}=10$ ns, a secret bit rate of $\sim70$ kHz over 1000 km is possible with a repeater station spacing of 2.6 km, a detection efficiency of $\eta_d=95\%$, a re-encoding error of $\epsilon_r=0.3\permil$, and using $[4,14,4]$-trees of 285 photons. For more modest gate times of $\tau_{\text{CZ}}=100$~ns, a secret bit rate of $\sim13$ kHz over 1000 km for the same parameters can be achieved (see above and supplemental material~\cite{SM} for a justification of these numbers for a concrete physical realization).

\subsection{Logical errors}
We have assumed a generic re-encoding error $\epsilon_r$ in our optimization above. This encoding error will, in general, be determined by errors from both the generation of the photonic tree-clusters and the re-encoding step. In the optimization, we assumed a fixed $\epsilon_r$ and optimized the tree-encoding for the given distances (see supplemental material~\cite{SM}). One could imagine that $\epsilon_r$ will depend on the size of the encoding. To investigate this, we consider single qubit depolarizing channels of the form in Eq.~(\ref{eq:depol}) acting on all qubits in the encoding. We can then estimate the single qubit error probability $\epsilon$ that will result in a given re-encoding error-probability, $\epsilon_{r}$ for a specific tree-encoding as detailed in the supplemental material~\cite{SM}. The tree encodings are remarkably robust to errors even in the presence of loss and we find that $\epsilon_r/\epsilon\approx3$ for tree-encodings and loss corresponding to the optimization in Fig.~\ref{fig:figure5} except for the high error ($\epsilon_r=0.1\%$) optimization where we find that $\epsilon_r/\epsilon\approx4-5$. Notably, we do not find any significant dependence of $\epsilon_r$ on the different tree-encodings. This is consistent with the errors of the two qubits (1st level qubit and root qubit) participating in the Bell measurement dominating the re-encoding error.

\subsection{Comparison to other approaches}

The proposed repeater compares favorably to previously proposed one-way repeater protocols~\cite{Muralidharan2014,Azuma2015,Ewert2017,Lee2018,Glaudell2016} (see Tab.1 in the supplemental material~\cite{SM}). It enables similar secret key rates for roughly the same error parameters and detection efficiencies. The key advantage of this repeater, however, is that it requires substantially less resources per repeater station than any of the previous protocols. In particular, the number of spin qubits per repeater stations is two orders of magnitude lower than the matter based protocol in Ref.~\cite{Muralidharan2014} and the large overhead of single photon sources for linear optics protocols~\cite{Azuma2015} is circumvented. The resources for the latter may be reduced by generating the photon cluster states as proposed in Ref.~\cite{Buterakos2017}. Nonetheless, the size of the encoding is still more than an order of magnitude larger than for our protocol. We obtain this reduction by directly encoding the message qubit in a tree-cluster state and using the spin-photon interface for near-deterministic re-encoding operations as opposed to swapping entanglement with probabilistic linear optics Bell measurements. Compared to other one-way repeaters, the proposed repeater is, however, not fault tolerant and the tolerable error level therefore decreases with the distance. This is simply a consequence of the buildup of error with each re-encoding operation and an order of magnitude decrease in error, thus roughly corresponds to an order of magnitude increase in achievable distance (e.g distances in the range $10^{3}-10^{4}$ km would be achievable with an error rate of $10^{-4}-10^{-5}$ and tree-cluster states of depth 3). It might be possible to remedy this effect by incorporating error correcting for logical errors at the expense of a few additional spins at each station. A full investigation of this is, however, beyond the scope of this article. 
 
We have also compared the repeater to a two-way quantum repeater allowing for parallel entanglement generation attempts using the same total number of spin qubits as our one-way repeater (see supplemental material~\cite{SM}). We have assumed deterministic noise-free entanglement swapping and noise-free entanglement generation using a two-photon interference scheme~\cite{Duan2003}. This provides a crude comparison with standard two-way repeaters. As shown in Fig.~\ref{fig:figure5b}, our one-way repeater reaches key rates orders of magitude higher than the two-way repeater due to the higher local repetition rate. Furthermore this advantage is achieved without the need for long coherence times. For the two-way repeater, orders of magnitude larger coherence times (miliseconds to seconds) are required. Note that Ref.~\cite{Muralidharan2016} contains an extended comparison between one-way and two-way repeaters also showing the advantage of the former in the low noise limit. The comparison performed so far has been made in terms of communication rate per qubit by requiring the two repeater approaches to have the same total number of qubits. In practice other factors may also be relevant for the comparison. In particular, the number of repeater stations is about an order of magnitude larger for the one-way repeater and the initial cost of establishing a quantum repeater chain is thus likely to be higher with the present approach. This is, however, compensated by a much higher communication rate resulting in a better rate per qubit. 
 
\section{Conclusion and Discussion}
We have proposed a novel one-way quantum repeater based on photonic tree-cluster states. The repeater enables secret bit rates $\sim 70$ kHz ($\sim$ 13 kHz) over a distance of 1000 km assuming GHz single photon emission rates and spin-spin entangling gate times of $10$ ns (100 ns). We have demonstrated how both the error correction and the generation of the tree-cluster states at the repeater stations can be performed with a minimum number of spin systems. Specifically, we have outlined a repeater setup that requires only a single quantum emitter and two memory spin qubits per station. As compared to the daunting requirement for realizing previously proposed one-way quantum repeaters, this places our proposal within experimental reach of current technologies.  

Solid state systems such as quantum dots and diamond defects are promising hardware candidates. Single photon emission rates exceeding GHz have already been achieved~\cite{Liu2018, Zhang2018} together with efficient coupling to nanophotonic waveguides and cavities~\cite{Sipahigil2016,Kalb2017,Delteil2015,De-Santis2017,Javadi2018}. The spin-spin gates required for the repeater may be mediated through tunneling in quantum dots~\cite{Bayer2001,Kim2010,Tian2014} or nuclear-electron spin coupling for diamond defects~\cite{Kalb2017}. While many of the key elements necessary for this proposal have already been demonstrated, additional engineering of the platforms will be required in order to reach the required photon collection efficiencies and error level of the gates. Importantly, the proposed implementations based on state-of-the-art solid state emitters appear capable of reaching those demanding metrics. Notably both high cooperativity and combined detection and in/out coupling efficiencies above 90\% have recently been reported with a SiV defect center coupled to a nanophotonic cavity~\cite{Mihir2019}. For quantum dots, chip-to-fiber coupling efficiencies exceeding 80\% have been reported~\cite{Daveau2016}, which could readily be improved further. High cooperativity has also recently been demonstrated by coupling a quantum dot to a microcavity~\cite{Najer2019}. The current state-of-the-art is thus not far from the required performance of this protocol. In addition, our protocol outperforms direct transmission already at overall detection efficiencies of $85\%$ and for slightly higher efficiencies, error levels $>0.1\%$ can be tolerated as shown in the supplemental material~\cite{SM}. Efficient frequency conversion to the telecom C-band where low loss optical fibers exist and fast optical switching will be necessary to achieve long communication distances. While this remains a challenge, there has been impressive progress on high-efficiency frequency conversion~\cite{Kambs2016,Dreau2018,Bock2018} and fast optical switching~\cite{Vedovato2018} with clear routes towards further improving the performance. An alternative strategy is to develop quantum dots emitting directly in the telecom C-band~\cite{Nawrath2019}, which relies on the continuous development of material growth technology. It thus seems feasible that our proposal provides a promising experimental route towards high-rate quantum key distribution with potential for proof-of-principle experiments with current technology. 
	\begin{acknowledgments}
We would like to thank Dirk Englund, Matthias Christandl, Martin Hayhurst Appel, Ralf Riedinger, and Mihir Bhaskar  for many valuable discussions. PL and AS gratefully acknowledges financial support from the Danish National Research Foundation (Center of Excellence `Hy-Q', grant number DNRF139), the European Research Council (ERC Advanced Grant `SCALE'), and the European Union's Horizon 2020 research and innovation programme under grant agreement No 820445 and project name Quantum Internet Alliance.  
JB acknowledges financial support from VILLUM FONDEN via the QMATH Centre of Excellence (Grant No. 10059). T.S. was supported by the European Union's Horizon 2020 research and innovation program under the Marie Skodowska Curie grant agreement no. 753067 (OPHOCS) and the Federal Ministry of Education and Research of Germany (BMBF, project DiNOQuant13N14921). Work at Harvard was supported by NSF, CUA, NSF EFRI, ARL and Vannevar Bush Fellowship. 
	\end{acknowledgments}
	\clearpage
\newpage
\onecolumngrid
 \section*{Supplemental Material: One-way quantum repeater based on near-deterministic photon-emitter interfaces}


\setcounter{section}{0}
In this supplemental material, we present a detailed error analysis of the spin-photon CZ-gate used in the re-encoding step of the repeater (Sec.~\ref{sec:Bell}) and the details of the numerical optimization (Sec.~\ref{sec:num}). Furthermore, we outline the experimental architecture of the repeater stations (Sec.~\ref{sec:exp}), present a comparison with previously proposed one-way repeater schemes and a generic two-way repeater (Sec.~\ref{sec:comp}), and finally discuss the minimal requirements for beating direct transmission (Sec.~\ref{sec:direct}).  

\section{Bell measurement} \label{sec:Bell}
In this section, we describe the details behind the photon-spin interaction used for the re-encoding step at the repeater stations. As detailed in the main article, we consider an emitter with ground states $\ket{0}_s$ and $\ket{1}_s$. The state $\ket{1}_s$ is coupled to an excited state $\ket{e}_s$ through the cavity field. In a frame rotating with the cavity resonance frequency, the interaction between the emitter and the cavity field is described by the Hamiltonian
\begin{equation}
 \hat{H}=g\ketbra{e}{1}_s\hat{c}+g^*\ketbra{e}{1}_s\hat{c}^{\dagger}, 
 \end{equation}
 where $\hat{c}$ is the annihilation operator of the cavity field, $g$ is the single photon Rabi frequency and we have assumed the transition $\ket{1}_s\leftrightarrow\ket{e}_s$ to be resonant with the cavity frequency. We assume that spontaneous emission from the excited level is described by a Lindblad operator $L_{\gamma}=\sqrt{\gamma}\ketbra{1}{e}$. The input-output relations of the cavity field are
 \begin{eqnarray}
 \dot{\hat{c}}&=&-i\left[\hat{c},\hat{H}\right]-\frac{\kappa}{2}\hat{c}+\sqrt{\kappa_{\text{in}}}\hat{a}_{\text{in}}+\sqrt{\kappa_{\text{loss}}}\hat{F}_{\kappa} \\
 \hat{a}_{\text{out}}&=&\hat{a}_{\text{in}}-\sqrt{\kappa_{\text{in}}}\hat{c}
 \end{eqnarray}
 where $\hat{a}_{\text{in}}$ ($\hat{a}_{\text{out}}$) are the annihilation operators for the input and output light of the cavity and we assume the input light to be resonant with the cavity. The total decay rate of the cavity field is $\kappa=\kappa_{\text{in}}+\kappa_{\text{loss}}$. The transmission of the input mirror is described by $\kappa_{\text{in}}$ while $\kappa_{\text{loss}}$ is the intra-cavity loss rate associated with vacuum noise operator $\hat{F}_{\kappa}$. Note that $\kappa_{\text{loss}}$ also includes loss due to imperfect mode-matching of an incoming photon to the cavity mode. In the single photon limit, the above equations can be solved using the Fourier transform assuming the emitter is initially in the ground state manifold. This gives the following expression for the output field~\cite{Anders2003}
\begin{eqnarray}
\hat{a}_{\text{out}}&=&\frac{1-2\kappa_{\text{in}}/\kappa+4C\hat{N}_1}{1+4C\hat{N}_1}\hat{a}_{\text{in}}\text{      + \emph{noise}}, \\
&=&s(\hat{N_1})\hat{a}_{\text{in}}\text{      + \emph{noise}},
\end{eqnarray}
where $C=\frac{\abs{g}^2}{\kappa\gamma}$ is the cooperativity of the system and $s(\hat{N_1})=\frac{1-2\kappa_{\text{in}}/\kappa+4C\hat{N}_1}{1+4C\hat{N}_1}$. We have collected the vacuum noise terms originating from intra-cavity loss and spontaneous emission in the term (+ \emph{noise}). The quantity $\hat{N}_{1}=\ketbra{1}{1}_s$ is the projector onto state $\ket{1}_s$. Thus, $\hat{N}_1=1$ if the emitter is prepared in state $\ket{1}_s$ and $\hat{a}_{\text{out}}\approx\hat{a}_{\text{in}}$ for $C\gg1$. If the atom is prepared in state $\ket{0}_s$, we have that $\hat{N}_1=0$ and $\hat{a}_{\text{out}}\approx-\hat{a}_{\text{in}}$ for negligible cavity loss ($\kappa\approx\kappa_{\text{in}}$). Consequently, the field experiences a $\pi$-phase shift depending on the atomic state.  

 To quantify the performance of the above two-qubit operation, we consider the state transfer of a photonic input qubit $\alpha\ket{0}+\beta\ket{1}$ in the time-bin encoding to a spin system initialized in $\ket{0}_s$ as needed for the re-encoding step in the repeater protocol. As described in the main article, the photonic qubit state $\ket{0}$ is first scattered off the cavity. Subsequently a $x$ rotation ($\ket{0}_s\to(\ket{0}_s+\ket{1}_s)/\sqrt{2}$) is performed on the spin system (assumed to have negligible error) before the photonic state $\ket{1}$ is scattered off the cavity. This leads to the following transformation
\begin{eqnarray}
(\alpha\ket{0}+\beta\ket{1})\ket{0}_s\to\frac{1}{\sqrt{2}}\Big[\alpha\ket{0}(s(0)\ket{0}_s+s(0)\ket{1}_s)+\beta \ket{1}(s(1)\ket{1}_s+s(0)\ket{0}_s)\Big]\text{ + \emph{vac}},
\end{eqnarray}
The state transfer is conditioned on detecting the photon (in the $x$-basis), which projects out the vacuum component resulting from intra-cavity loss and spontaneous emission from the emitter. We can therefore consider the heralded state
\begin{equation}
\ket{\psi}=\sqrt{\frac{\eta_d}{2P_s}}\Big[\alpha \ket{0}(s(0)\ket{0}_s+s(0)\ket{1}_s)+\beta\ket{1} (s(1)\ket{1}_s+s(0)\ket{0}_s)\Big],
\end{equation}
where $P_s=\eta_d(\abs{\alpha}^2\abs{s(0)}^2+\abs{\beta}^2(\abs{s(1)}^2+\abs{s(0)}^2)/2)$ is the success probability of the operation assuming photon detection efficiency of $\eta_d$.  We can quantify the performance of the operation by calculating the fidelity between $\ket{\psi}$ and the state resulting from the perfect operation, i.e. $F=\abs{\braket{\psi}{\psi_{\text{per}}}}^2$, where
\begin{equation}
\ket{\psi_{\text{per}}}=\frac{1}{\sqrt{2}}(-\alpha\ket{0}(\ket{0}_s+\ket{1}_s)+\beta\ket{1}(\ket{1}_s-\ket{0}_s)). 
\end{equation}  
In the limit $C\gg1$ and $\kappa_{\text{loss}}/\kappa_{\text{in}}\ll1$, we can expand the fidelity to get
\begin{equation}
F\approx1-(1+\abs{\alpha}^2)\abs{\beta}^2\left(\left(\frac{\kappa_{\text{loss}}}{\kappa_{\text{in}}}\right)^2-\frac{\kappa_{\text{loss}}}{2\kappa_{\text{in}}C}+\frac{1}{16 C^2}\right),
\end{equation}
while the success probability will be 
\begin{equation}
P_s\approx\eta_d\left(1-2(1+\abs{\alpha}^2)\frac{\kappa_{\text{loss}}}{\kappa_{\text{in}}}-\abs{\beta}^2\frac{1}{2C}\right).
\end{equation}
It is seen that having $C\sim\kappa_{\text{in}}/\kappa_{\text{loss}}\sim100$ is enough to ensure that $1-F\lesssim10^{-4}$ and $P_s\approx 0.99\eta_d$. 

So far, we have assumed that the frequency width of the photon ($\sigma_{\text{ph}}$) is narrow enough to neglect errors from the finite bandwidth ($\sim C\gamma$) of the Purcell enhanced emitter. The error due to finite bandwidth will be suppressed at least as $(\sigma_{\text{ph}}/(C\gamma))^2$~\cite{sahand2016,Witthaut2012}. Consequently, having $\sigma_{\text{ph}}\sim\gamma$ and $C\sim100$ is enough to ensure errors on the order of $10^{-4}$. Finally, reflection of photons into the detector due to e.g. imperfect mode-matching or from another optical component in the setup can also lead to errors. Such reflections correspond to an operation with no spin-dependent phase accumulation. Consequently, the probability of a faulty reflection directly maps into the error probability of the operation. Careful engineering of the mode profile and additional filtering of undesired modes (e.g. in polarization) can be employed to suppress back reflections to the desired level. Modelling imperfect mode-matching as a fictitious beam splitter before the resonator, however, reveals that the probability of back reflections only appears as the mode-matching inefficiency squared. Thus, a mode-matching efficiency of 99\% is sufficient to ensure an error of $10^{-4}$. Note that we are referring to the mode-matching efficiency of the full system. Reflections from any component in the system other than the bare cavity basically correspond to an effective cavity with slightly different resonance conditions.

\section{Numerical optimization} \label{sec:num}

In this section, we present the details of the numerical optimization of the repeater including the simulation of how single qubit depolarizing errors add to a total re-encoding error. 

\begin{figure} [h]
\centering
\subfloat{\label{fig:figureS3a}\includegraphics[width=0.48\textwidth]{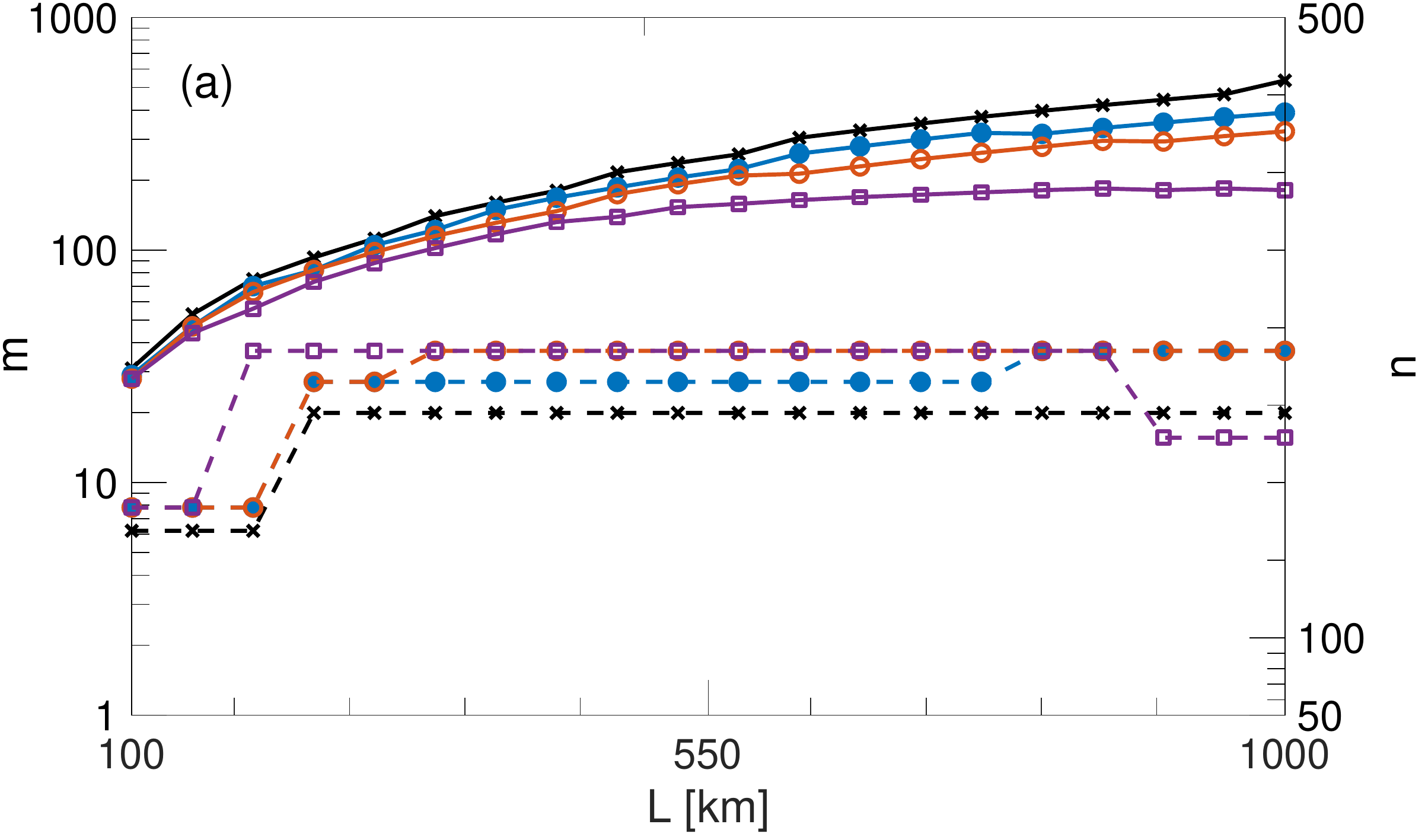}}
\subfloat{\label{fig:figureS3b}\includegraphics[width=0.48\textwidth]{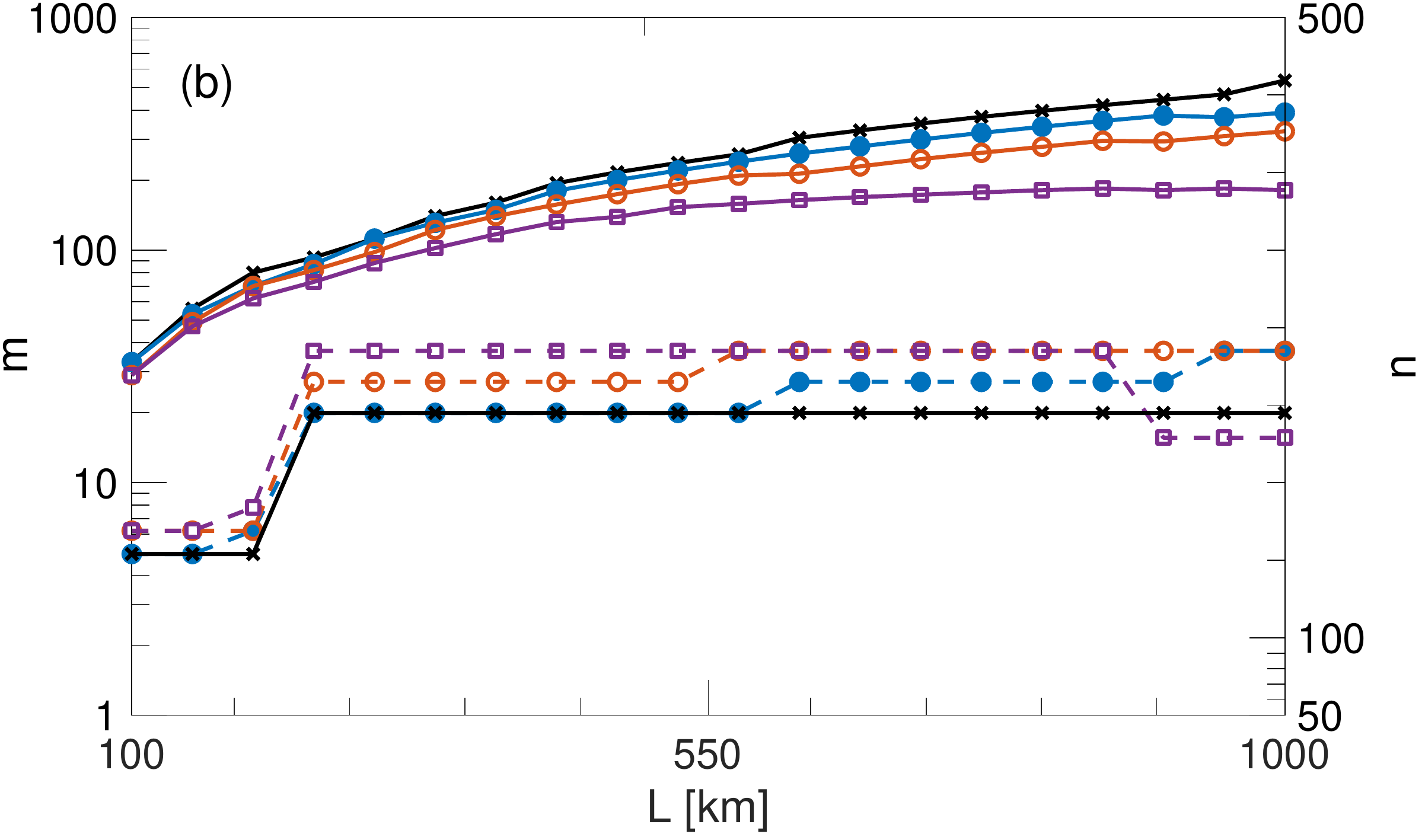}}
\caption{Optimized number of repeater stations $m$ (solid lines, left axis) and total number photons in the tree-encoding $n$ (dashed lines, right axis) corresponding to the results shown in Fig.~6 of the main article. The parameters for spin-spin gate times of $10\tau_{\text{ph}}$ are shown in (a) while the parameters for gate times of $100\tau_{\text{ph}}$ are shown in (b). The markers in the plots correspond to: $\bm{\times}:\epsilon_r=0.1\permil$, $\textcolor{dblue}{\bullet}:\epsilon_r=0.3\permil$, $\textcolor{orange}{\bm{\circ}}:\epsilon_r=0.5\permil$, and $\textcolor{purple2}{\bm{\square}}:\epsilon_r=0.1\%$. Here $\epsilon_r$ is the re-encoding error as defined in the main article.}
\label{fig:figureS3}
\end{figure}

\begin{figure} [h]
\centering
\subfloat{\label{fig:figureS3a}\includegraphics[width=0.48\textwidth]{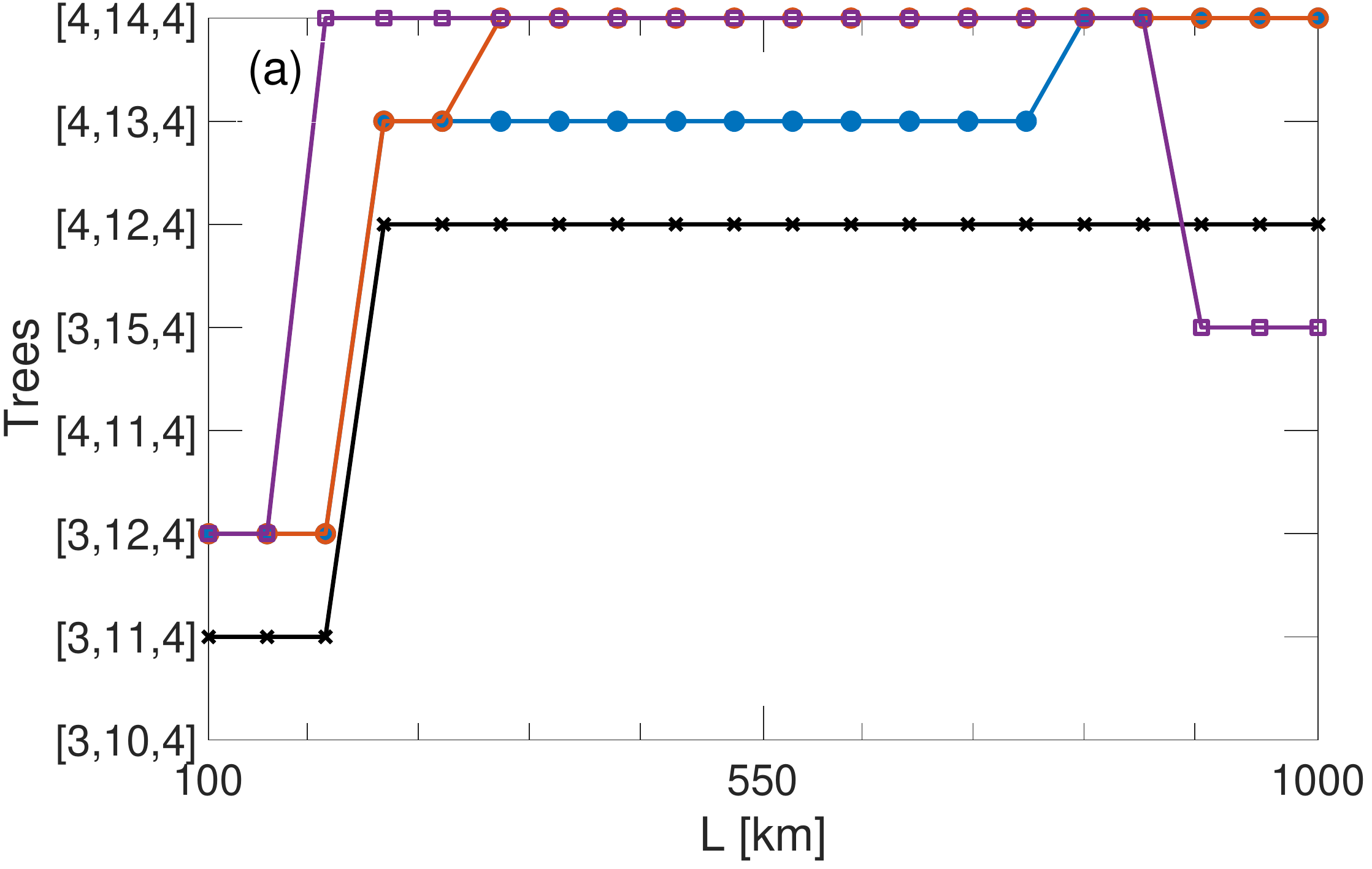}}
\subfloat{\label{fig:figureS3b}\includegraphics[width=0.48\textwidth]{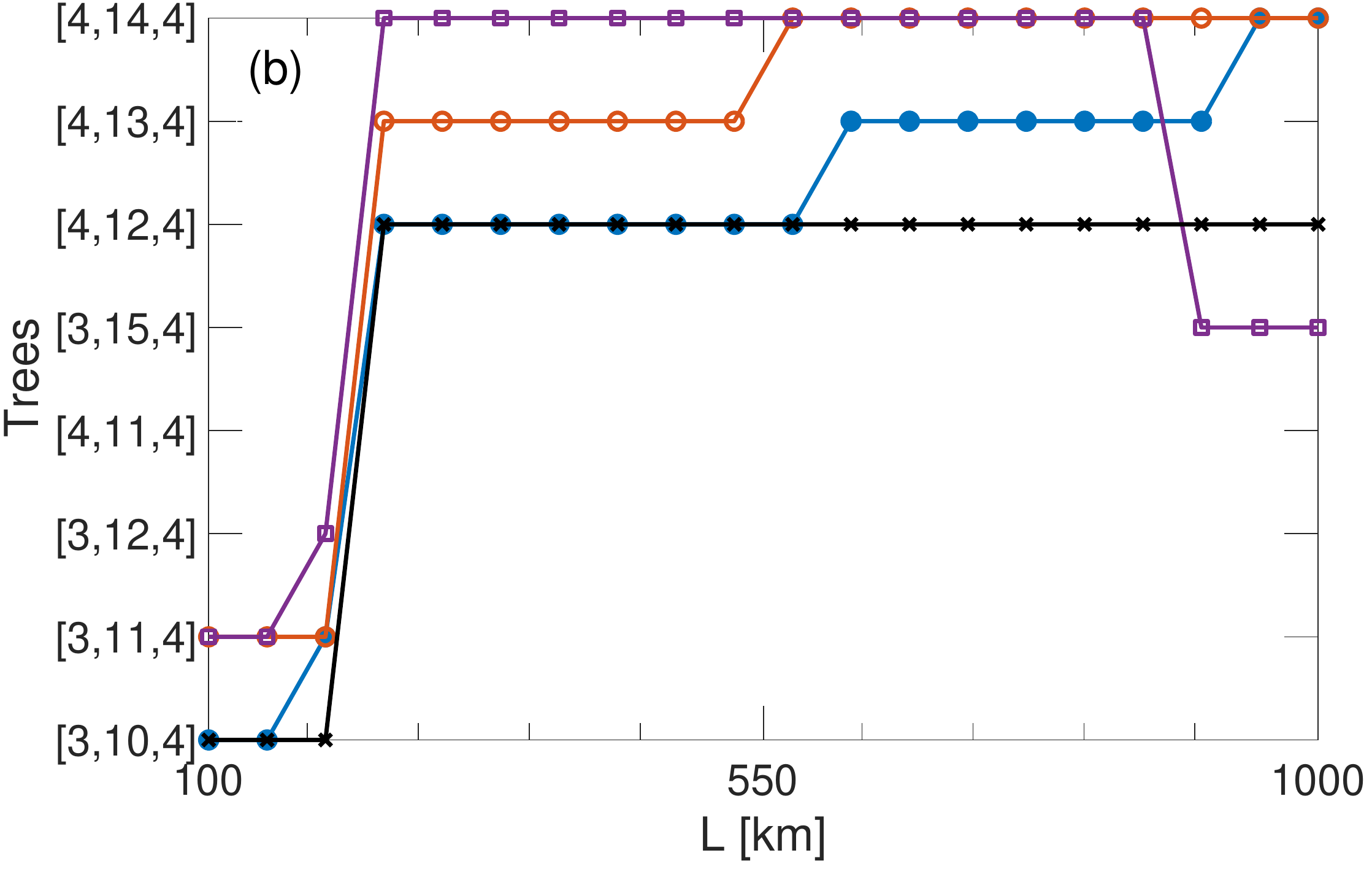}}
\caption{Optimal tree encodings corresponding to the results shown in Fig.~6 of the main article. The encodings for spin-spin gate times of $10\tau_{\text{ph}}$ are shown in (a) while the same for gate times of $100\tau_{\text{ph}}$ are shown in (b). The markers in the plots correspond to: $\bm{\times}:\epsilon_r=0.1\permil$, $\textcolor{dblue}{\bullet}:\epsilon_r=0.3\permil$, $\textcolor{orange}{\bm{\circ}}:\epsilon_r=0.5\permil$, and $\textcolor{purple2}{\bm{\square}}:\epsilon_r=0.1\%$. Here $\epsilon_r$ is the re-encoding error as defined in the main article.}
\label{fig:figureS4}
\end{figure}

The number of repeater stations ($m$) and total number of photons ($n$) in the encodings that correspond to the optimized performance in Fig.~6 of the main text is shown in Fig.~\ref{fig:figureS3}. The corresponding tree cluster states are shown in Fig.~\ref{fig:figureS4}. Note that the optimization is allowed to use more encoding levels, but always find optimal tree encodings of depth 3. 

\subsection{Single qubit errors}
We wish to investigate how depolarizing errors on individual qubits of the tree-cluster encoding adds to a total re-encoding error, $\epsilon_r$. To this end,  we consider single qubit depolarizing channels of the form 
\begin{equation} \label{eq:depol}
\Lambda_{i}(\hat{\rho})=(1-\epsilon_i)\hat{\rho} + \frac{\epsilon_i}{3}\left(\sigma^x_i\hat{\rho}\sigma^x_i+\sigma^y_i\hat{\rho}\sigma^y_i+\sigma^z_i\hat{\rho}\sigma^z_i\right)
\end{equation}
acting on the $i$'th qubit of the encoding. Here $\hat{\rho}$ describes a multi-qubit tree-cluster state and $\sigma^{x,y,z}_i$ are the Pauli matrices. While the error-parameters $\epsilon_i$ might be different in general, we will consider them to be equal ($\epsilon_i=\epsilon$ for all $i$) for simplicity. Assuming the above depolarizing channels, we can investigate how the single qubit depolarizing errors ($\epsilon$) adds to a total re-encoding error-probability, $\epsilon_{r}$ for a specific tree-encoding. 
 
\begin{figure} [H]
\centering
\subfloat{\label{fig:figureS5a}\includegraphics[width=0.48\textwidth]{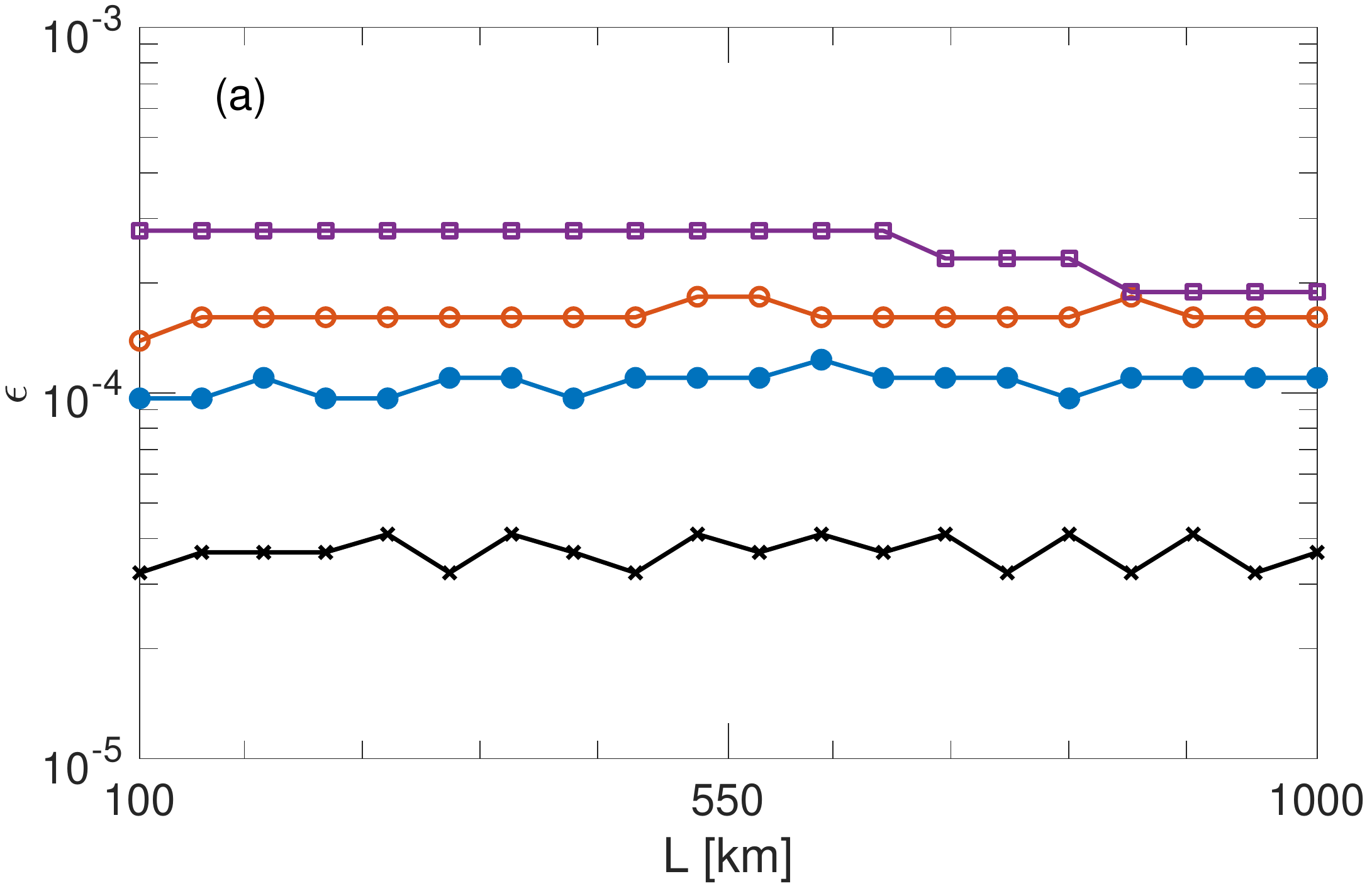}}
\subfloat{\label{fig:figureS5b}\includegraphics[width=0.48\textwidth]{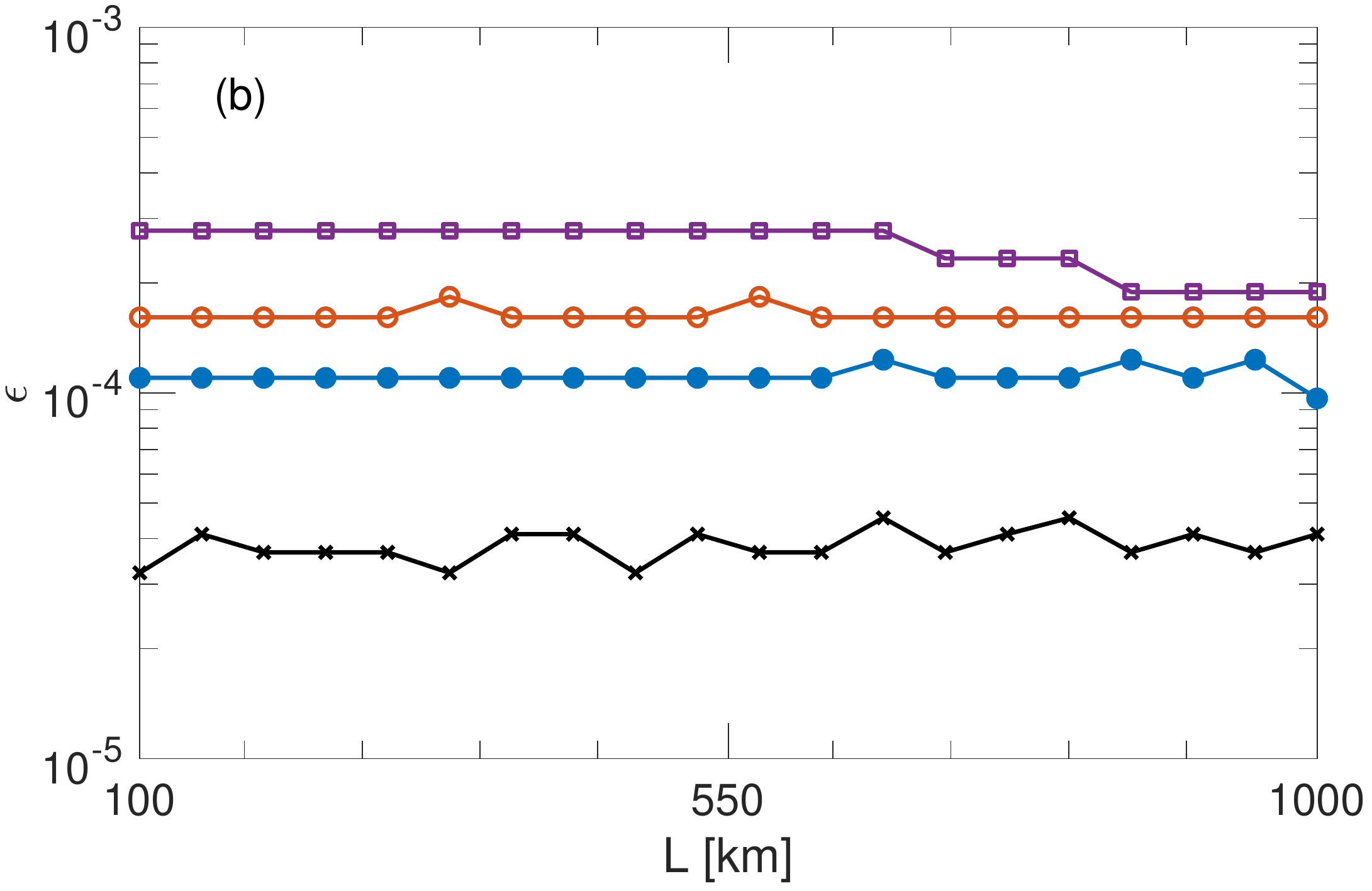}}
\caption{Maximum tolerable single qubit depolarizing error $\epsilon$ corresponding to re-encoding errors of $\bm{\times}:\epsilon_r=0.1\permil$, $\textcolor{dblue}{\bullet}:\epsilon_r=0.3\permil$, $\textcolor{orange}{\bm{\circ}}:\epsilon_r=0.5\permil$, and $\textcolor{purple2}{\bm{\square}}:\epsilon_r=0.1\%$. The simulations were performed with tree-encodings and loss corresponding to the optimization in Fig.~6 in the main article with spin-spin gate times of $10\tau_{\text{ph}}$ in (a) and $100\tau_{\text{ph}}$ in (b). The data points were obtained through averaging of $10^6$ simulations per point.}
\label{fig:figureS5}
	\end{figure}

We determine $\epsilon_r$ through numerical simulations where Pauli matrices are randomly applied to every qubit of the encoded tree and the root qubit of the new tree as dictated by Eq.~(\ref{eq:depol}). All qubits of the encoded tree-cluster that are not lost are measured and a majority vote is performed to determine the correct value of the necessary $z$-measurements for the re-encoding step. For example, if three qubits are measured that should all correspond to the same inferred value of a $z$-measurement on a fourth qubit, a correct measurement occurs if at most one of these have an error. The result of such simulations are shown in Fig.~\ref{fig:figureS5}.  

\section{Experimental implementation} \label{sec:exp}
In this section, we outline the requirements for photonic tree-cluster state generation and outline the design of repeater stations.  

The time budget for the generation of the tree states is determined by the time of photonic qubit generation and the time needed to perform spin-spin gates. In the generation scheme, each branch of the tree is emitted sequentially. For a tree with branching vector $\vec{t}=[b_0,b_1,b_2]$, the time it takes to emit one branch is estimated as
\begin{equation}
\tau_{\text{branch}}\approx(100+(1+b_2)b_1)\tau_{\text{ph}}+(b_1+3)\tau_{\text{CZ}},
\end{equation}
where $\tau_{\text{ph}}$ is the generation time of a photonic time-bin qubit and $\tau_{\text{CZ}}$ is the time of spin-spin CZ-gates. Here we have assumed that the first level photon has a generation time of $100\tau_{\text{ph}}$ in order to sufficiently suppress errors in the re-encoding operation due to finite bandwidth of the emitter (see Sec.~\ref{sec:Bell}). Furthermore, the creation of the first level photon requires 2 CZ gates, as outlined in the main text, and one CZ gate is required to entangle with the root qubit resulting in the factor of 3 in the expression. The total time it takes to generate a tree will be $\tau_{\text{tree}}\approx \tau_{\text{branch}}b_0$. Since the photons are emitted from the bottom up, a delay line is needed to ensure that the 1st-level qubits are measured first at the re-encoding step. This delay line has to be of length $l_{\text{del}}\approx\tau_{\text{branch}}c$, where $c$ is the speed of light in the delay fibre. To route the higher-level qubits into the delay line, fast optical switching has to be employed. Assuming a Purcell-enhanced photon emission lifetime time of about 100~ps~\cite{Liu2018,Zhang2018}, and a total delay time of 500~ps for each emission event during which the excitation has decayed with more than 99\% probability, gives a total generation time of $\tau_{\text{ph}}\sim1$~ns for the two subsequent excitation events required to generate a time-bin qubit. Thus, optical switching rates of more than 2~GHz are required. 

For the spin-spin gate, we consider both fast timescales on the order of $\tau_{\text{CZ}}\sim10$~ns~\cite{Kim2010,Evans2018} and more modest gate times of $100$ ns. For these example parameters, the largest $[4,14,4]$-tree found in the optimization (see Fig.~\ref{fig:figureS3}b) can be emitted in time $\tau_{\text{tree}}\sim1$ $\mu$s ($\tau_{\text{branch}}\sim 340$ ns) assuming $\tau_{\text{CZ}}\sim10$~ns or $\tau_{\text{tree}}\sim7$ $\mu$s ($\tau_{\text{branch}}\sim 1870$ ns) assuming $\tau_{\text{CZ}}\sim100$~ns. Consequently, assuming $c=2\cdot10^{8}$ m/s,  delay lines of length $l_{\text{del}}\sim68$ m or $l_{\text{del}}\sim374$ m are needed for the bottom level qubits of the tree to ensure the right detection order.  

\begin{figure}[h]
\centering
\includegraphics[width=1\textwidth]{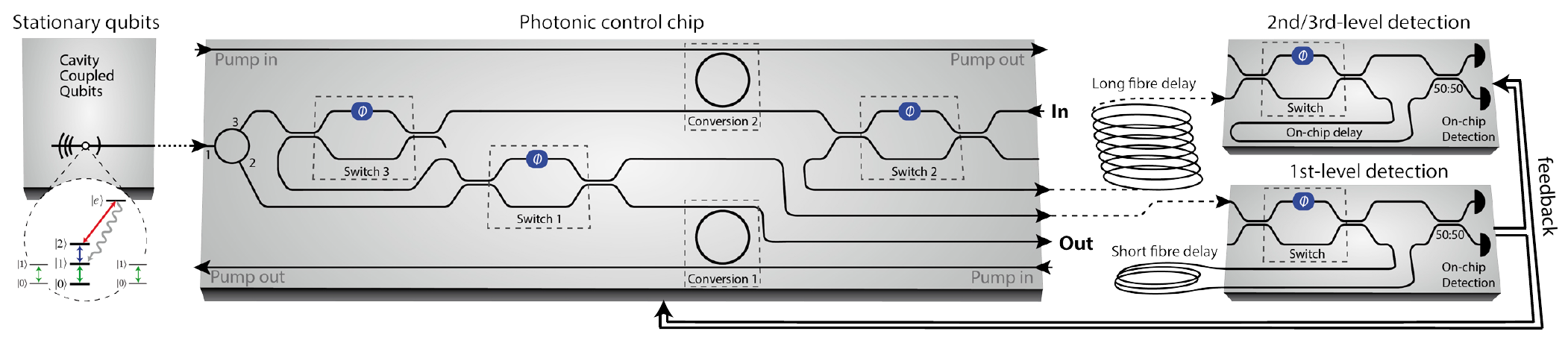}
\caption{Illustration of a physical unit to implement cluster state generation and re-encoding. Similar to the detection unit (Fig. 5 in the main article), we propose an architecture of on-chip photonic circuits that is extended by the required matter qubits, which are efficiently coupled to a one-sided cavity. To fulfill the specific requirements of each section of the quantum repeater unit we propose a modular chip architecture. On-chip Faraday circulators and integrated Mach-Zehnder interferometers with electro-optic modulation provide the required routing and fast switching. For the generation of tree cluster states (left chip), a system of quantum emitter and two quantum memory spins can be used where the emitter couples to photon states and the other two qubits serve as root and ancilla qubits (see inset). Importantly, only the emitter qubit needs to be optically coupled to the outgoing photonic qubits. In the generation step, photons are sent directly to the photonic control chip (middle chip) where they are converted to the telecom C-band and directed to the next repeater station via the output port. Gates between the spins can be performed through tunneling (quantum dots) or electron-nuclear interaction (NV, SIV). For the re-encoding, the 1st-level qubits are converted to the emitter qubit wavelength, interact with the cavity coupled emitter and are then measured in the 1st-level detection unit (bottom right chip). If a previous 1st-level qubit has been detected (successful storage), the subsequent 1st-level qubits are sent directly to detection without interacting with the emitter. The 2nd- and 3rd-level qubits are not converted back from the telecom band but are sent directly to the 2nd/3rd-level detection unit to be measured either in the z- or x-basis depending on success or no-success (photon loss) of the measurement of the 1st level photons. The delay line in front of the telecom detection unit (top right chip) rearranges the order of the incoming tree qubits such that the 1st-level qubits are detected before the lower level qubits of their branches arrive. The switching of the subsequent 1st level qubits and measurement basis of the lower level qubits is the only feedforward required for the protocol. Since the long fiber delay is hundreds of ns, this sets the time scale of the feedback (faster switching is required, but once a detection is registered or not the measurement sequence is determined for the entire branch).  
}
\label{fig:schemeI}
\end{figure}

To achieve the required optical switching rates of more than 2~GHz we propose, similarly to the detection unit, an electro-optic photonic circuit. Hybrid integration schemes can be applied to efficiently couple photonic elements in the host material of the matter qubits to on-chip photonic circuits in optical modulation compatible materials~\cite{mouradian2015,murray2015,davanco2017}. For a repeater station, we propose such a hybrid photonic platform that consists of several individual photonic chips to tailor each photonic integrated cirtcuit to the specific requirements of a the station. More specifically, the proposed architecture consists of a chip that hosts the stationary qubits required for the cluster state generation; a chip to enable photonic routing, switching and frequency conversion; a chip for the detection of 1st-level cluster state qubits; and a chip for the detection of 2nd/3rd-level qubits. A sketch of a repeater station is shown in Fig.~\ref{fig:schemeI}.

\section{Comparison with other quantum repeater schemes} \label{sec:comp}

\subsection{One-way quantum repeaters}
We have compared our proposed one-way quantum repeater protocol with a number of previously proposed one-way schemes~\cite{Muralidharan2014,Azuma2015,Ewert2017,Lee2018,Glaudell2016}. While the architectures are very different, Table~\ref{tab:tableS1} provides a high-level comparison of rates for specific error levels and outlines the general requirements of the different architectures. However, we stress that the overall feasibility of e.g. all-optical and matter based architectures cannot readily be compared due to the pronounced differences of the hardware. 

As noted in the main article, the current repeater enables similar secret key rates as the previous protocols but requires substantially less resources per repeater station. Specifically, Ref.~\cite{Muralidharan2014} requires two orders of magnitude more spin systems per repeater station while Ref.~\cite{Azuma2015} requires more than an order of magnitude larger encoding than our protocol due to the probabilistic operation of linear optics Bell measurements for entanglement swapping. Importantly, the encoding considered in Refs.~\cite{Ewert2017,Lee2018} allows for efficient linear optics Bell measurements on the encoded level. While this circumvents the bottleneck of linear optics Bell measurements in the re-encoding process, the probabilistic operation remains a challenge for the generation of the logical states. Ref.~\cite{Lee2018} does not provide any estimate of the resources needed for state generation while Ref.~ \cite{Ewert2017} considers the possibility of generating the photonic states with optical nonlinearities arriving at 3 orders of magnitude more optical nonlinearities compared to the number of spin systems in our proposal. The type of nonlinearity considered in Ref.~\cite{Ewert2017} does not make use of the repeated emission possible with a quantum emitter. It is, however, an open question how to generate the parity encoding with repeated emission from quantum emitters and it is beyond the scope of this work to address this question. One immediate issue is that in order to be compatible with linear optics Bell measurements, simultaneous arrival of photons is required. The protocol in Ref.~\cite{Glaudell2016} has very small repeater station spacing resulting in almost two orders of magnitude more repeater stations and requires quantum qudits. 

\begingroup
\begin{table*} [t]
\begin{tabular}{|p{1.5 cm}|p{7 cm}|p{6 cm}|p{3 cm}|}
\hline
Scheme & Characteristics & Performance & Physical/logical qubits \\ \hline
Current repeater &  One-way repeater employing tree-encoding to battle transmission loss. Re-encoding requires a single succesful Bell measurement independent of the encoding size. Requires a small amount of feedforward.  & Rate at $1000$ km: $r_{s}t_{0}\approx0.1$ for $\eta_d=0.95,\epsilon=10^{-4}, L_{\text{att}}=20$ km, and $L_{0}=2.6$ km. & 285 photons per logical qubit. One quantum emitter and 2 auxiliary spin qubits per repeater station.   \\ \hline   
Ref.~\cite{Muralidharan2014} & Based on the quantum parity code and teleportation based error correction with matter qubits. Number of matter qubits and CNOT gates used for re-encoding scales linearly with encoding size. &  Optimal rate at $1000$ km: $r_{s}t_{0}\approx 0.6$ for $\eta_d=0.9, \epsilon=10^{-4}, L_{\text{att}}=20$ km, and $L_{0}=1.6$ km. & 126 photons per logical qubit and 252 matter qubits per repeater station.   \\ \hline
Ref.~\cite{Azuma2015} & Based on tree-clusters as photonic memories. Multiple tree-encoded qubits are generated at repeater nodes with linear optics requiring feedforward.  & Rate at $1000$ km: $r_{s}t_{0}\approx 0.6$ for $\eta_d=0.95, \epsilon=4.2\cdot10^{-5}, L_{\text{att}}=22$ km, and $L_{0}=4$ km. Optimally, cluster states of 19 logical qubits per repeater station are generated corresponding to a total of $4864$ photonic qubits. & Cluster state of 4864 photons per repeater station requiring $\gtrsim10^6$ single photon sources per station~\cite{Pant2017} .    \\ \hline
Ref.~\cite{Ewert2017} & Based on the quantum parity code with linear optics implementation. Number of Bell measurements for re-encoding scales linearly with size of the encoding. Generation of encoded states envisioned both with linear optics and optical nonlinearities. No feedforward required  & Rate at $1000$ km: $r_{s}t_{0}\approx 0.8$ for $\eta_d=0.97, \epsilon=1\cdot10^{-3}, L_{\text{att}}=22$ km, and $L_{0}=1.3$ km.& 737 photons per logical qubit requiring 1473 optical nonlinearities or $\gtrsim10^6$ single photon sources per repeater station. \\ \hline
Ref.~\cite{Lee2018} & Based on the quantum parity code but considers linear optics implementation. Number of Bell measurements for re-encoding scales linearly with size of the encoding. Generation of encoded states is envisioned with linear optics but detailed resource analysis is not provided. Requires feedforward.& Rate at $1000$ km: $r_{s}t_{0}\approx 0.7$ for $\eta_d=0.95, \epsilon=5.6\cdot10^{-5}, L_{\text{att}}=22$ km, and $L_{0}=1.8$ km.  & 464 photons per logical qubit. \\ \hline
Ref.~\cite{Glaudell2016} & Considers CSS codes and operation in a sequential manner. Implementation with minimum number of matter qudits considered.  & Estimated near MHz rates for $L=100-1000$ km assuming $\eta_d\approx1,\epsilon\approx10^{-5}, L_{0}=50$ m and loss in telecom fibers (not further specified). Photon emission rates of $\sim10$ GHz were considered. & 7 photons per logical qubit and $4$ matter qudits per repeater station.  \\ \hline
\end{tabular}
\caption{Performance and general characteristics of quantum repeaters considered in Refs.~\cite{Muralidharan2014,Azuma2015,Lee2018,Glaudell2016}. Here, $r_s$ is the secret bit rate, $L_0$ is the spacing between repeater stations, $\epsilon$ is the single qubit error probability and $L_{\text{att}}$ is the attenuation length in fibers. $t_0$ is the time of state generation and error correction at the repeater stations. The generalized detection efficiency $\eta_d$ may include single photon source efficiency, in/out coupling losses and single photon detection efficiency depending on the specific scheme. 
}
\label{tab:tableS1}
\end{table*}
\endgroup

\subsection{A generic two-way repeater}
To benchmark the proposed one-way quantum repeater against two-way quantum repeaters, we consider a generic two-way protocol. We assume perfect entanglement swapping and error-free entanglement generation in the elementary links using a two-photon interference protocol~\cite{Duan2003} with a middle station. As a result, no entanglement purification is needed, which boosts the rate. The tree-clusters in the tree-repeater are generated using 2 memory spins and one quantum emitter per repeater station so the total number of spin qubits used in the tree repeater for a given distance is $3(m+1)$. We allow these resources (qubits) to be equally distributed between the $l$ repeater stations in the two-way repeater for parallel entanglement attempts where $l$ is an odd number. The secret key rate of the two-way repeater is then estimated as
\begin{equation}
r=\frac{1}{Z_{l}(p_{\text{ent}})}\times\frac{(l+1)c}{L},
\end{equation}
where $Z_{l}(p)=\sum_{k=1}^{l+1}\binom{l+1}{k}\frac{(-1)^{k+1}}{1-(1-p)^k}$ accounts for the number of tries to generate entanglement in $l+1$ pairs where each pair succeeds with probability $p$~\cite{Bernardes2011}. The success probability of entanglement generation in a link is estimated as~\cite{Borregaard2015}.
\begin{equation}
p_{\text{ent}}\approx1-(1-\frac{1}{2}\eta_d^2e^{-\frac{L}{(l+1)L_{\text{att}}}})^{\lceil(3(m+1))/(2(l+1))\rceil},
\end{equation}
which is the probability that at least one entangled pair is generated using $\lceil(3(m+1))/(2(l+1))\rceil$ parallel attempts. We use the ceiling function $\lceil\quad\rceil$ to get an upper bound. In the comparison, we use the number of repeater stations from the tree-repeater with re-encoding error $\epsilon_r=0.1\permil$. The secret key rate is optimized over the number of repeater stations $l$ to take into account the trade-off between having faster entanglement generation  due to short separation between repeater stations and due to multiple parallel channels. The result of the optimization is included in Fig.~5 of the main text, and is significantly below the rate of our proposed one-way repeater.  

Another notable difference between two-way repeaters and one-way repeaters is the difference in repeater node spacing.  Two-way repeaters operate, in general, with an order of magnitude longer repeater node spacing than one-way repeaters, which may be more compatible with existing infrastructure. Nonetheless, parameters such as the total cost of repeater nodes may be lower for the present approach if the same performance is required. As the above comparison shows, many more qubits per repeater node are needed for a two-way repeater to reach comparable rates to the one-way repeater. Consequently, e.g. the number of cryostats per repeater node will be substantially higher for a two-way repeater which may be more costly than the higher number of potentially cheaper repeater stations in the one-way repeater.     	

\section{Beating direct transmission} \label{sec:direct}

As detailed in the main text, errors on the order of $0.1\permil$ and detection efficiencies around $0.95\%$ are required for fast long distance quantum communication with one-way repeaters. It is, however, possible to outperform direct transmission with the proposed quantum repeater protocol for significantly relaxed parameters. 

\begin{figure} [H]
\centering
\subfloat{\includegraphics[width=0.48\textwidth]{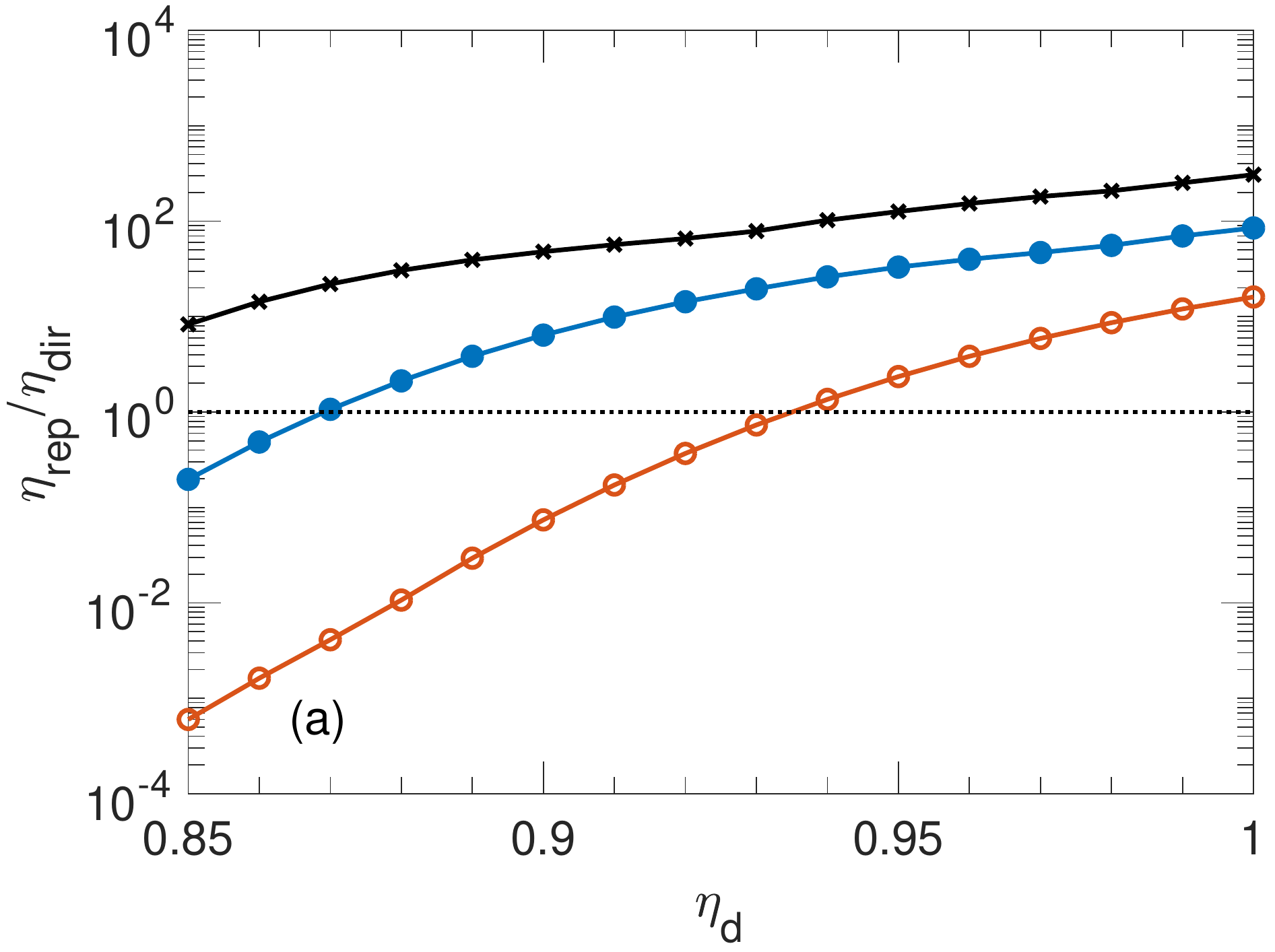}} 
\subfloat{\includegraphics[width=0.45\textwidth]{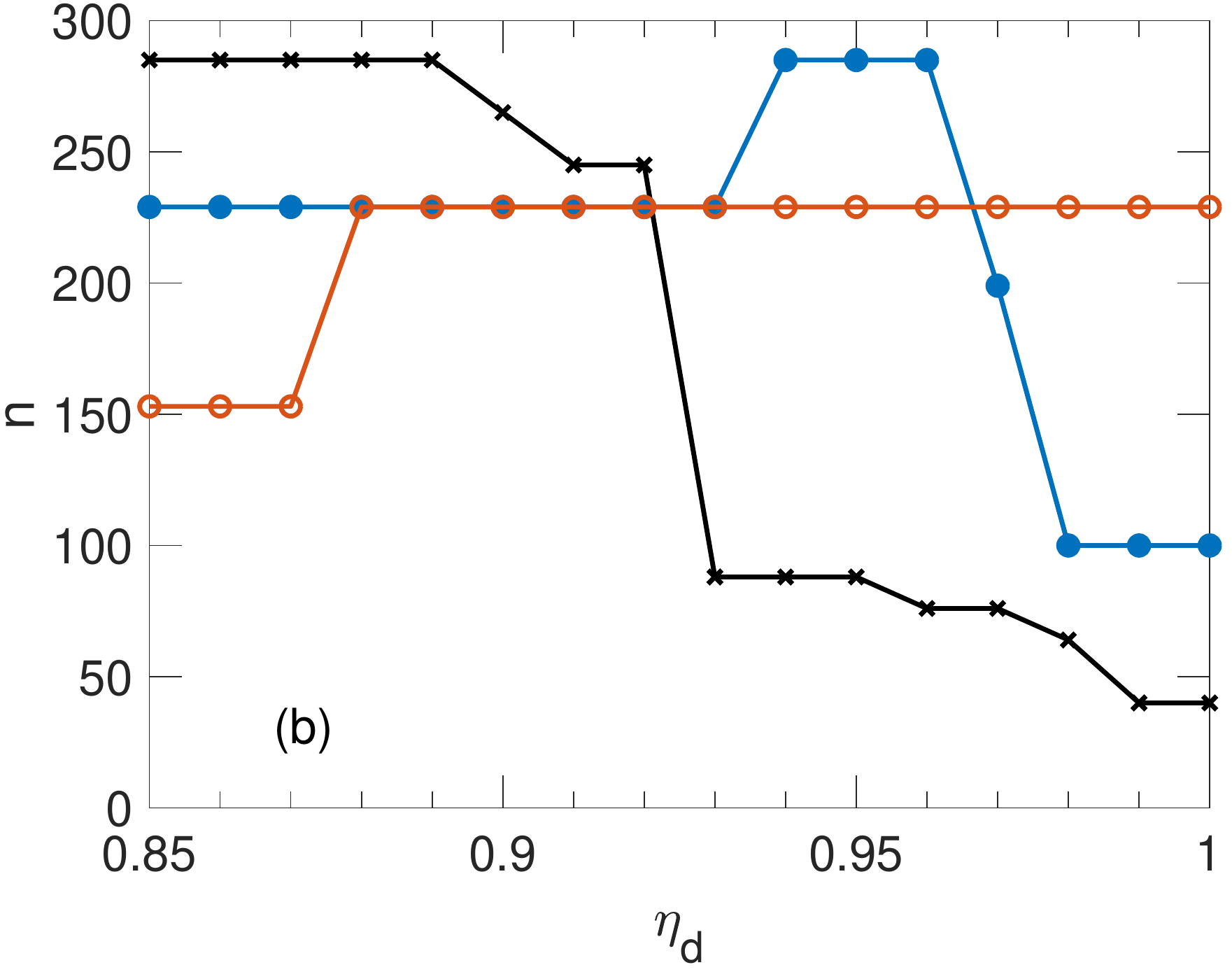}}
\caption{(a) Maximum ratio ($\eta_{\text{rep}}/\eta_{\text{dir}}$) between the transmission of the tree-repeater and direct transmission as a function of the detection efficiency $\eta_d$ for a distance of 200 km. $\times:\epsilon_r=0.1\%$, $\textcolor{dblue}{\bullet}:\epsilon_r=0.2\%$, and $\textcolor{orange}{\circ}:\epsilon_r=0.3\%$. The number of photons in the corresponding optimal tree-encodings ($n)$ are shown in (b). We have restricted the minimization to trees with $n\leq300$ photons and the optimal number of repeater stations was found to be 97 for $\epsilon_r=0.1\%$, 49 for  $\epsilon_r=0.2\%$, and 32 for  $\epsilon_r=0.3\%$ for all values of $\eta_d$ in the plots. }
\label{fig:figureB1}
\end{figure}

We consider direct transmission as a scenario where single photons are used to transmit a qubit directly from the start station to the end station. If $n$ single photons are emitted, at least one of them has to arrive at the end station to have a succesful transmission. The probability for this is
\begin{equation}
\eta_{\text{dir}}=1-(1-\eta_de^{-L/L_{\text{att}}})^n,
\end{equation}      
where $\eta_d$ is the detection efficiency, $L$ is the total distance, and $L_{\text{att}}=20$ km is the attenuation length of optical fibers as in the main text. 

We can compare this transmission probability to the transmission of a tree-repeater. For a fair comparison, we fix the number of photons in the direct transmission $n$ to the number of photons in the tree-encoding i.e. the size of the encoding. As detailed in the main text, the transmission probability of the tree-repeater is $\eta_{\text{rep}}=\eta_{e}^{m+1}$ where $m=L/L_{0}-1$ is the number of repeater stations between the start and end station and $\eta_e$ is the effective transmission of the encoded qubit information between repeater stations. We include the possibility of re-encoding errors at the repeater stations with probability $\epsilon_r$ and require that the total error probability of the transmitted qubit ($\epsilon_{\text{trans}}\approx(m+1)\epsilon_{r}$) is at most 10\%.  
To compare the tree-repeater with direct transmission, we find the number of repeater stations and the tree-encoding that maximize $\eta_{\text{rep}}/\eta_{\text{dir}}$ for given $\eta_d$ and $\epsilon_r$ over a distance of $L=200$ km. The result of this optimization is shown in Fig.~\ref{fig:figureB1}. It is seen that the tree-repeater outperforms direct transmission already for $\eta_d=85\%$ and a re-encoding error of $\epsilon_r=0.1\%$ using an encoding of $n=285$ photons and $m=97$ repeater stations for optimal performance. Increasing the detection efficiency both increase the rate of the tree-repeater and decreases the encoding size to below 100 photons for $\eta_d\gtrsim0.93$. Higher detection efficiency also enables the tree-repeater to beat direct transmission for larger re-encoding errors of $\epsilon_r=0.2\%$ and $0.3\%$ where respectively, $m=49$ and $32$ repeater stations are used for optimal performance. The optimal tree-encodings all have depth 3.

	\clearpage

\end{document}